\documentclass[5p]{elsarticle}
\usepackage[usenames,dvipsnames]{color}
\usepackage[normalem]{ulem}
\usepackage{amsmath}
\usepackage{todonotes}
\usepackage{listings}
\usepackage[hyphens]{url}
\usepackage[switch]{lineno} 
\usepackage{microtype}
\usepackage{xcolor}
\usepackage[binary-units]{siunitx}
\usepackage{subcaption}
\usepackage[T1]{fontenc}
\usepackage{lmodern}
\usepackage{hyperref}
\usepackage{textcomp}

\usepackage{cleveref}
\crefname{lstlisting}{listing}{listing}

\def\backtick{\char18}
\lstdefinestyle{mystyle}{literate={`}{\backtick}1, escapechar=@}
    
\usepackage{booktabs}
\usepackage{tabularx}

\makeatletter
\providecommand{\doi}[1]{doi:#1}
\makeatother
\biboptions{compress}

\usepackage[utf8]{inputenc}
\usepackage[english]{babel}
\usepackage[frozencache]{minted}
\usepackage{xcolor}
\definecolor{LightGray}{gray}{0.9}

\usepackage[acronym,shortcuts]{glossaries}
\newacronym{api}{API}{Application Programming Interface}
\newacronym[longplural={bounding volume hierarchies}]{bvh}{BVH}{bounding volume hierarchy}
\newacronym{rdf}{RDF}{radial distribution function}
\newacronym[longplural={potentials of mean force and torque}]{pmft}{PMFT}{potential of mean force and torque}
\newacronym{tbb}{TBB}{Intel Threading Building Blocks}
\newacronym{pmf}{PMF}{potential of mean force}
\newacronym{bood}{BOOD}{bond-orientational order diagram}
\newacronym{pypi}{PyPI}{the Python Package Index}
\newacronym{cna}{CNA}{Common Neighbor Analysis}
\newacronym{fft}{FFT}{fast Fourier transform}
\newacronym{msd}{MSD}{mean squared displacement}

\newcommand{\freud}{\texttt{freud}}
\newcommand{\compute}{\texttt{compute}}

\newcommand{\nq}{\texttt{NeighborQuery}}
\newcommand{\nlist}{\texttt{NeighborList}}
\newcommand{\system}{\texttt{system}}
\newcommand{\neighbors}{\texttt{neighbors}}
\newcommand{\aabb}{\texttt{AABBQuery}}
\newcommand{\lc}{\texttt{LinkCell}}
\newcommand{\cit}[1][]{%
    \textbf{Add Citation for 
    \ifthenelse{\equal{#1}{}}{}{: #1}}%
}
\newcommand\dchi[0]{\Delta\xi_{12}}
\newcommand\ftilde[0]{\Tilde{F}_{12}}

\usepackage{mathtools}
\DeclarePairedDelimiter\norm{\lVert}{\rVert}%

\title{\freud: A Software Suite for High Throughput Analysis of Particle Simulation Data}
\author[umichche]{Vyas~Ramasubramani}
\ead{vramasub@umich.edu}

\author[umichphy]{Bradley~D.~Dice}
\ead{bdice@umich.edu}

\author[umichmse]{Eric~S.~Harper}
\ead{harperic@umich.edu}

\author[umichche]{Matthew~P.~Spellings}
\ead{mspells@umich.edu}

\author[umichche]{Joshua~A.~Anderson}
\ead{joaander@umich.edu}

\author[umichche,umichphy,umichmse,umichbi]{Sharon~C.~Glotzer\corref{cor1}}
\ead{sglotzer@umich.edu}

\cortext[cor1]{Corresponding author\\ \textcopyright This manuscript version is made available under the CC-BY-NC-ND 4.0 license \url{http://creativecommons.org/licenses/by-nc-nd/4.0/}.}

\address[umichche]{Department of Chemical Engineering, University of Michigan, Ann Arbor, MI 48109}
\address[umichphy]{Department of Physics, University of Michigan, Ann Arbor, MI 48109}
\address[umichmse]{Department of Materials Science and Engineering, University of Michigan, Ann Arbor, MI 48109}
\address[umichbi]{Biointerfaces Institute, University of Michigan, Ann Arbor, MI 48109}

\begin{document}

\begin{abstract}
The \freud\ Python package is a library for analyzing simulation data.
Written with modern simulation and data analysis workflows in mind, \freud\ provides a Python interface to fast, parallelized C++ routines that run efficiently on laptops, workstations, and supercomputing clusters.
The package provides the core tools for finding particle neighbors in periodic systems, and offers a uniform \glstext{api} to a wide variety of methods implemented using these tools.
As such, \freud\ users can access standard methods such as the radial distribution function as well as newer, more specialized methods such as the \acrlong{pmft} and local crystal environment analysis with equal ease.
Rather than providing its own trajectory data structure, \freud\ operates either directly on NumPy arrays or on trajectory data structures provided by other Python packages.
This design allows \freud\ to transparently interface with many trajectory file formats by leveraging the file parsing abilities of other trajectory management tools.
By remaining agnostic to its data source, \freud\ is suitable for analyzing any particle simulation, regardless of the original data representation or simulation method.
When used for on-the-fly analysis in conjunction with scriptable simulation software such as HOOMD-blue, \freud\ enables smart simulations that adapt to the current state of the system, allowing users to study phenomena such as nucleation and growth.
\end{abstract}


\maketitle

\noindent
{\bf PROGRAM SUMMARY}

\noindent
{\em Program Title:} \freud \\
{\em Licensing provisions:} BSD 3-Clause \\
{\em Programming language:} Python, C++ \\
{\em Nature of problem:}
Simulations of coarse-grained, nano-scale, and colloidal particle systems typically require analyses specialized to a particular system.
Certain more standardized techniques -- including correlation functions, order parameters, and clustering -- are computationally intensive tasks that must be carefully implemented to scale to the larger systems common in modern simulations.
\\
{\em Solution method:}
\freud\ performs a wide variety of particle system analyses, offering a Python API that interfaces with many other tools in computational molecular sciences via NumPy array inputs and outputs.
The algorithms in \freud\ leverage parallelized C++ to scale to large systems and enable real-time analysis.
The library's broad set of features encode few assumptions compared to other analysis packages, enabling analysis of a broader class of data ranging from biomolecular simulations to colloidal experiments.
\\
{\em Unusual features:}
\begin{enumerate}
    \item \freud\ provides very fast, parallel implementations of standard analysis methods like RDFs and correlation functions.
    \item \freud\ includes the reference implementation for the \gls{pmft}.
    \item \freud\ provides various novel methods for characterizing particle environments, including the calculation of descriptors useful for machine learning.
\end{enumerate}
{\em Additional comments:}
The source code is hosted on GitHub (\url{https://github.com/glotzerlab/freud}), and documentation is available online (\url{https://freud.readthedocs.io/}).
The package may be installed via \texttt{pip install freud-analysis} or \texttt{conda install -c conda-forge freud}.
\\



\glsresetall

\section{Introduction}
\label{sec:Introduction}

Molecular simulation is a crucial pillar in the investigation of scientific phenomena.
Increased computational resources, better algorithms, and new hardware architectures have made it possible to simulate complex systems over longer timescales than ever before \cite{Anderson2017b,Simon2019,Niethammer2014,Freddolino2008a,Shaw2009}.
The sheer volume of data necessitates computationally efficient analysis tools, while the diversity of data requires flexible tools that can be adapted for specific systems.
Additionally, to support scientists with limited prior computing experience, tools must be usable without extensive knowledge of the underlying code.

Numerous software packages that satisfy these requirements have been developed in recent years.
Tools such as MDTraj \cite{McGibbon2015}, MDAnalysis \cite{Michaud-Agrawal2011}, LOOS \cite{Romo2009}, MMTK \cite{Hinsen2000}, and VMD \cite{HUMP96} provide efficient implementations of various standard analysis methods.
Although powerful, such tools are generally limited in scope to all-atom simulations, particularly biomolecular simulations.
This focus is manifested not only through the features these tools provide, but also in their general design philosophies.

Perhaps the most pronounced characteristic of such tools is a strong emphasis on trajectory management, which includes parsing trajectory files and supporting extensive topology selection features to enable, for instance, selecting all residues or atoms in a protein backbone.
Although such tools are crucial for working with topologies in atomistic simulations, they are frequently cumbersome for working with coarse-grained simulation data where the trivial selection (all particles in the system) is the most common selection for various analyses.
Moreover, such topology selection tools make assumptions that are inappropriate for non-atomistic systems: ``bonding'' in colloidal systems, for instance, is typically based on whether two particles are found to be in the same neighborhood by some distance-based metric, not by the presence of a true chemical bond.
Since such determination of nearest neighbors is highly dynamic and parameter-dependent, it must be calculated on-the-fly and cannot be stored in a trajectory.

Another inconvenient but almost universal implementation choice is to directly tie analysis methods to trajectories by writing code that acts directly on some in-memory representation of a trajectory.
This direct linkage is generally inflexible because it inhibits pre-processing of the data before running the analysis, which is often crucial to analyzing more specialized systems.
More importantly, existing tools emphasize implementations of highly specific analyses involving, for instance, hydrogen bonding and protein secondary structure (using, e.g., DSSP \cite{Kabsch1983}), which are far less useful for analyzing non-biomolecular systems.
The predominant analyses of coarse-grained, colloidal-scale, or nanoparticle simulations usually involve measurements like numbers of nearest neighbors, diffraction patterns, or bond-orientational order parameters.
These analyses bear little relation to the analyses performed for atomistic systems.
These considerations suggest a need for a different type of analysis package that offers different methods than most existing tools.

\begin{figure*}[t!]
    \centering
    \includegraphics[width=0.8\textwidth]{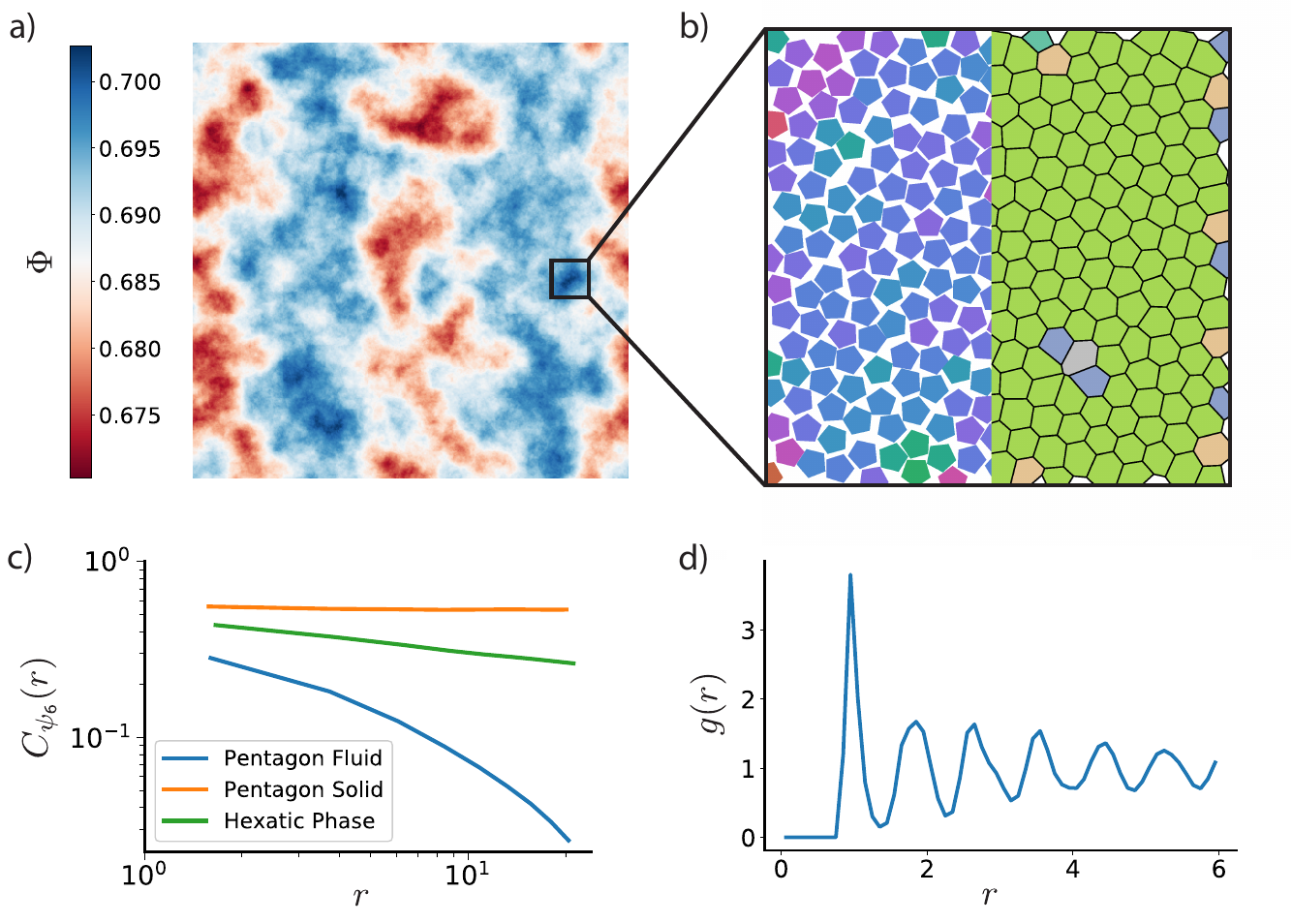}
    \caption{The \freud\ library is capable of computing a number of characteristics of a system of particles. Here, we demonstrate some of those features on a 2D Monte Carlo simulation of polygons that exhibits hexatic ordering \cite{Anderson2017b}. a) Phase separation is clearly evident in this system of $512^2$ pentagons colored by local density $\Phi$; the system is divided into denser (blue) and less dense (red) regions. b) Zooming into a particularly dense region shows that the hexatic ordering (left half) is generally uniform across the region. The Voronoi diagram of the system (right half) is also largely defect-free, with just a few pentagons having more or fewer than 6 nearest neighbors. c) The spatial correlation of the hexatic order parameter $C_{\psi_6}(r)$ is nearly constant for a nearly perfect crystal of pentagons (orange), whereas it decays very quickly in a fluid (blue). For a comparable system of hexagons, however, we see a power-law decay (green) in the hexatic order parameter due to the presence of a hexatic phase between the solid and fluid phases. d) The radial distribution function $g(r)$ for the system of pentagons shows the expected sequence of neighbor shells as a function of distance.}
    \label{fig:Overview}
\end{figure*}

In this paper we introduce \freud, an open-source simulation analysis toolkit that addresses these needs.
All inputs to and outputs from \freud\ are numerical arrays of data, and the package makes no reference to predefined notions of atoms or molecules.
As a result, \freud\ can analyze particle-based data from both experiments and simulations regardless of the specific tools, methods, or software that were used to generate it.
The package provides a Python \gls{api} for accessing fast methods implemented in C++, and it implements numerous specific methods such as radial distribution functions and correlation functions that are common in the field of soft-matter physics (see \cref{fig:Overview}).
Prior works have used \texttt{freud} for:
determining spatial correlation functions and \glspl{pmft} in two dimensions \cite{Anderson2017b};
calculating Steinhardt order parameters for identifying solid-like particles \cite{Reinhart2018,Howard2018};
computing spherical harmonics for machine learning on crystal structures \cite{Spellings2018a};
optimizing pair potentials for designing complex crystals \cite{Adorf2018b};
calculating strain fields by finding neighbors of particles against a uniform grid \cite{VanSaders2018};
finding \glspl{pmft} in depletion-mediated self-assembly of hard cuboctahedra \cite{Karas2016a};
measuring rotational degrees of freedom in entropically ordered systems \cite{Antonaglia2018};
umbrella sampling of solid-solid phase transitions using Steinhardt order parameters \cite{Du2016a};
evaluating \glspl{pmft} in analysis of two-dimensional shape allophiles \cite{Harper2015a};
and more.
The \freud\ library is designed to work well with coarse-grained particle models, such as those used in simulations of anisotropic nanoparticles, colloidal crystals, and polymers, and its methods are particularly useful for studies of phase transitions and critical phenomena in such systems.
The package is likely to be of greatest interest to scientific communities in materials science, chemical engineering, and physics, though many of its analysis methods would be useful in generic particle systems.
The \freud\ library also integrates well into the scientific Python ecosystem, especially in data pipelines for machine learning and visualization \cite{Dice2019}.

The paper is organized as follows.
We first address the core design principles that went into building \freud\ in \cref{sec:Design}.
\Cref{sec:Implementation} focuses more specifically on the details of the code, including information on class structures.
\Cref{sec:Features} describes the various analysis methods in \freud\ and details their uses.
Finally, in \cref{sec:Examples} we provide some example code demonstrating the usage of \freud \footnote{The code for these examples and many others is available at \url{https://github.com/glotzerlab/freud-examples} and in our online documentation.}.
The figures in this paper are rendered using Matplotlib \cite{Hunter2007a} unless otherwise noted.

\section{Design}
\label{sec:Design}

Many of the best known tools for analyzing molecular simulations are built into either simulation toolkits (such as LAMMPS \cite{Plimpton1995a}, GROMACS \cite{Berendsen1995a}, or the cpptraj \cite{Roe2013} plugin to Amber \cite{Case2005ThePrograms}) or visualization toolkits (such as VMD \cite{HUMP96}, PyMOL \cite{PyMOL}, or OVITO \cite{Stukowski2010a}).
Although most of these have introduced varying degrees of scripting support over the years, the analyses built into simulation toolkits are primarily focused on performing one-shot analyses on trajectory files directly from the command line.
The visualization toolkits tend to have more full-featured scripting interfaces, but they are frequently difficult (if not impossible) to use outside their own sandboxed environments, complicating or even prohibiting integration with other tools.
More recently, many newer tools such as MDTraj \cite{McGibbon2015}, MDAnalysis \cite{Michaud-Agrawal2011}, LOOS \cite{Romo2009}, and Pteros \cite{Yesylevskyy2012} have aimed to decouple analysis from simulation and visualization, making scriptability a primary focus to increase flexibility.
Among such tools, Python is the most common language of choice due to its ease of use and the fact that it can be naturally extended with high performance languages like C, C++, and FORTRAN.

\freud\ follows in the footsteps of these tools, providing a full-featured Python \gls{api} to access all of its routines.
However, while most other tools focus on calculating properties involving molecular topologies, \freud\ is fundamentally designed for analyzing the local neighborhoods of particles, particularly where such local analyses provide global insight about the system.
Such analyses are typically far more varied and system-dependent than the standard analyses of molecular topologies and therefore require more flexible tools.
To meet this need, \freud\ eschews any form of trajectory object encoding system topology and is instead designed such that each analysis method is an independent Python class that performs computations directly on NumPy arrays \cite{Oliphant2006a} of data.

This design makes it possible to use a far wider range of data with \freud\ than is possible with tools that are tied to simulation trajectories.
For instance, calculating Voronoi diagrams and computing spatial correlation functions with \freud\ is possible for essentially arbitrary spatial data, not just the result of a molecular simulation.
Another major benefit is that NumPy arrays (the de facto standard for numerical data in Python) can be easily passed between multiple tools, making \freud\ equally easy to use for one-off analyses or as part of a larger analysis pipeline involving many steps and using various software packages.
As a result, \freud\ is a much more flexible choice both for analyzing disparate sources of data and for incorporating into Python workflows.
For example, most of \freud's analyses can be used within the OVITO visualization environment for real-time visualization with almost no noticeable performance cost.

Producing such array data from simulation trajectories for input to \freud\ is straightforward because high quality file parsers with Python \glspl{api} already exist for all common trajectory file formats.
Through integration with tools like MDAnalysis, GSD \cite{GlotzerLabGSD}, and garnett \cite{GlotzerLabGarnett}, \freud\ can be used with data from over 25 different file formats, including common formats like DCD, XTC, and TRJ files.
\freud\ integrates with the trajectory objects produced by many of these tools, but if necessary, users can read trajectories into arrays and modify them before passing the data to \freud\ for analysis.
By using data read by other tools, \freud's analyses can be made aware of molecular topology if needed, but only when the analysis method requires such information.
Similarly, since the outputs of \freud's analyses are also NumPy arrays, they can be passed to almost any tool in the scientific Python software stack.
For example, constructing a Pandas \cite{Mckinney2010DataPython} \texttt{DataFrame} from the outputs of any \freud\ analysis requires just one line of code, and it immediately enables writing the output to text, CSV, or HDF5, or saving into an SQL database.

Beyond differences in trajectory and data handling, the most significant design choice in \freud\ stems from the most common pattern followed by its many analysis methods.
Since the first task in characterizing local neighborhoods is often the identification of neighboring particles, \freud\ provides efficient methods for finding neighbors in arbitrary system geometries.
The nearest-neighbor finding routines are designed to be as fast and flexible as possible, supporting various algorithms optimized for different system configurations and offering different criteria for neighbor selection.
In general, queries can be based on either a cutoff distance or a desired number of nearest neighbors.
These tools are optimized to provide cheap access to neighbors even in highly performance-critical loops in C++ analysis routines, but the package's system box representation and neighbor finding tools also have Python \glspl{api}, so users can implement custom analyses directly in Python (for an example, see \cref{sec:cna}).

The analysis methods in \freud\ are essentially independent tools that make use of these objects to efficiently perform various calculations.
These features are all presented with a common \gls{api}, easing the transition between the different types of analyses needed for different simulations.
All methods in \freud\ are accelerated through extensive parallelization.

\section{Implementation}
\label{sec:Implementation}

The \freud\ package is entirely object-oriented, with two core C++ classes: the \texttt{Box} class, which encapsulates all logic associated with periodicity in arbitrary triclinic boxes (boxes with 3 linearly independent basis vectors); and the \nq\ class, which facilitates efficiently finding, storing, and iterating over nearest neighbors.
In keeping with the Python ethos, box objects in \freud\ may be constructed from a variety of inputs.
Any method in \freud\ that accepts a box object also accepts a number of objects that can be interpreted as a box, such as a $3\times 3$ NumPy array of box vectors or a list of three numbers representing the edge lengths of an orthorhombic box.
There are two subclasses of \nq\ in \freud\ that each implement different neighbor search algorithms: one implements a \gls{bvh} \cite{Anderson2016b}, while the other implements a cell list \cite{Allen1987}.
The \nlist\ class is a lightweight storage mechanism for \nq\ results that accelerates performing multiple analyses on the same set of neighbor pairs.

The analysis methods in \freud\ are encapsulated by \emph{Compute classes}, which are loosely defined as classes providing a \compute\ method that populates class attributes after performing some computation.
Compute classes, such as the \texttt{density} module's \texttt{RDF} class, are usually configured with constructor arguments, after which they can be used multiple times to perform distinct calculations.
Some classes in \freud\ (e.g. the \gls{rdf}, \gls{pmft}, or \acrlong{bood}) represent histogram-like quantities, and therefore allow the user to specify \texttt{reset=False} as an argument to \compute\ in order to accumulate and average data over many calls.

Compute classes can be divided into two groups, those that depend on finding neighbors and those that do not.
A majority of calculations in \freud\ require neighbors, and the \compute\ methods of such classes all share two arguments, \system\ and \neighbors\ (in addition to analysis-specific arguments like particle orientations for \glspl{pmft}; such arguments are also typically NumPy arrays).
The \system\ parameter accepts a \nq\ or any object that can be interpreted as a tuple \texttt{(box, points)}, where the \texttt{box} is any valid box-like object (as described above) and the \texttt{points} argument is anything that can be interpreted as an $N\times3$ NumPy array of positions.
Valid systems include simulation frame objects from tools such as MDAnalysis, GSD, garnett, OVITO, or the particle simulation engine HOOMD-blue \cite{Anderson2008d,Glaser2015g,Anderson2016b}.

\begin{figure*}[ht!]
    \centering
    \includegraphics[width=0.9\textwidth]{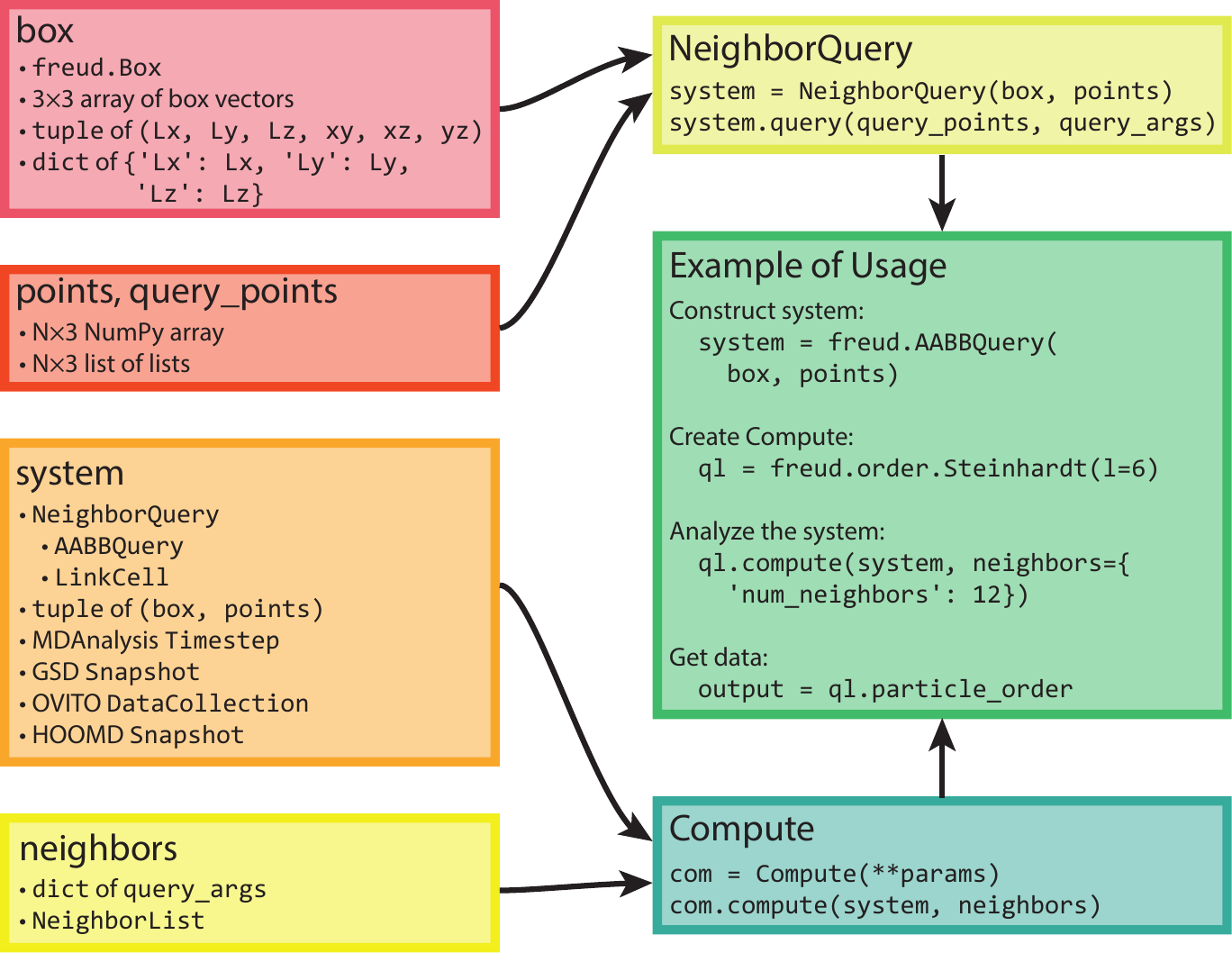}
    \caption{Here we show the flow of various types of inputs into \freud.
    Boxes can be constructed based on a variety of inputs, all of which can also directly be provided anywhere a \texttt{box} object is required.
    Similarly, any object that can be interpreted as an $N\times3$ array can be provided where particle positions are required.
    Any valid pair of box and points can be used to construct a \nq\ object, which is one of the types of \system s that \freud\ accepts.
    In addition to a \nq, \freud\ can also interpret raw tuples of boxes and points as \system\ objects, or use simulation frames from numerous external tools (a subset are shown in the figure).
    Any computation that involves finding nearest neighbors also requires a specification of \neighbors, which can be a \nlist\ or a dictionary of query arguments.
    The Example of Usage on the right shows a typical use case of \freud\ that combines these concepts.
    }
    \label{fig:DataFlow}
\end{figure*}

When performance is critical, providing a \nq\ object is advantageous because many \compute\ methods can reuse these neighbor search data structures.
For all other \system\ inputs, \freud\ internally constructs a \nq\ if the \compute\ method requires neighbor pairs.
The \neighbors\ argument is a dictionary of query arguments, such as \texttt{dict(num\_neighbors=12)} or \texttt{dict(\\r\_max=3.0)} (the complete specification for \freud's Query API is provided in the documentation).
Alternatively, users may precompute a \nlist\ and provide it as the \neighbors.
In this case, whether \system\ is a \nq\ or not has no impact on performance because the calculation will be carried out directly on the provided set of neighbor pairs and no additional spatial searches are required.
\Cref{fig:DataFlow} shows a flowchart demonstrating how these classes and data structures are used.

Some methods in \freud\ do not operate on neighboring pairs of particles.
For those that still depend on particle positions (such as \texttt{GaussianDensity}), the first argument is still any valid \texttt{system}, but no \texttt{neighbors} are provided.
Some methods do not depend on positions at all; for instance, the \texttt{Nematic} order parameter only requires particle orientations.
In such cases, the user can simply pass that quantity alone to the calculation.
This mode of operation is particularly useful when performing high-throughput analysis of large files; using file formats like GSD that permit reading only certain properties of the trajectory, users can minimize I/O operations by only reading the required arrays from memory.

All Compute classes use efficient, thread-parallel C++ implementations for performance-critical components.
The Python bindings for these C++ classes are generated using Cython \cite{Behnel2011a}, and the C++ methods are mirrored in Python using thin Cython classes that dispatch calls to the underlying C++ class instances.
The Cython classes have limited responsibilities: managing the memory of the underlying C++ instances, sanitizing inputs when necessary, and providing transparent access via memory views on C++ arrays.

The main exception to this design is the \texttt{msd} module, which is implemented in pure Python in \freud.
The \ac{msd} is a measure of, on average, how far particles move in a given window of time.
In a simulation trajectory of $N_f$ frames, the \ac{msd} of particle $i$ over a window of length $m$ frames is given by:
\begin{linenomath*}
    \begin{equation}
         MSD(i, m) = \frac{1}{N_f-m} \sum_{k=0}^{N_f-m-1} \norm{(\vec{r}_i(k+m) - \vec{r}_i(k))}^2
    \end{equation}
\end{linenomath*}
Therefore, the total \ac{msd} is given by:
\begin{linenomath*}
    \begin{equation}
         MSD(m) = \frac{1}{N_{p}} \sum_{i=1}^{N_{p}} MSD(i, m)
    \end{equation}
\end{linenomath*}
Direct computation of the \ac{msd} is an $\mathcal{O}(N_p N_f^2)$ operation, but by using a \ac{fft} this cost can be reduced to $\mathcal{O}(N_p N_f \log(N_f))$ \cite{Calandrini2011}.
When using this approach, the \ac{fft} is responsible for most of the computation time, and since packages like NumPy \cite{Oliphant2006a} and SciPy \cite{Jones2001} already expose fast C and FORTRAN \gls{fft} routines to Python, \freud\ simply leverages them directly and implements the rest of the \ac{msd} in pure Python.

Calculations in \freud\ are generally parallelized over \emph{particles} (e.g. the \texttt{Nematic} order parameter class) or over \emph{pairs of particles} (e.g. computing inter-particle distances for an \gls{rdf} with the \texttt{RDF} class).
Both the number of particles and the number of particle-particle pairs increase with system size, ensuring that the work is load-balanced well among threads because the number of threads is much less than the number of particles or pairs.
Parallelism in \freud\ is accomplished using \gls{tbb} \cite{Intel2018}.
Analysis routines are written as lambda functions operating on a particle or a pair of particles; \freud\ provides wrappers that then automatically parallelize these functions appropriately using \gls{tbb}.
Modern compilers aggressively inline such lambda functions, thereby optimizing away any additional cost that could arise from the extra function calls.
\freud\ uses thread-local storage extensively to avoid any parallel writes to data containers.
For histograms that accumulate over many frames of simulation data, \freud\ performs reduction over thread-local containers lazily.

Currently, \freud\ is at version 2.1.0 and supports Python versions 3.5.0 or greater.
The package is distributed through \gls{pypi} and the \texttt{conda-forge} channel of the Anaconda package manager \cite{2020AnacondaDistribution}, making it easy to install on any Unix-based operating system (e.g. Linux or macOS).
Builds for the Windows operating system are also available on \texttt{conda-forge}.
\freud\ depends on NumPy and TBB libraries, which are automatically installed with \freud.
The library can also be compiled from source using a C++11 compliant compiler.
Compilation requires NumPy and TBB headers as well as a Cython installation.
Code documentation is written using Google-style docstrings rendered using Sphinx and hosted on ReadTheDocs.
The \freud\ library is released open source under the BSD 3-Clause License, and the source code is available in a GitHub repository \cite{GlotzerLabFreud}.
Continuous integration testing is performed using CircleCI.

\section{Features}
\label{sec:Features}

\subsection{General Utilities}
The general utilities in \freud\ are contained in two modules: \texttt{box} and \texttt{locality}.
The \texttt{box} module contains the core \texttt{Box} class.
The \texttt{locality} module contains the \nq\ abstract class, which defines the standardized query \gls{api}.
\nq\ results (neighboring particle pairs) can be obtained dynamically or stored in the \nlist\ class provided by the \texttt{locality} module.

Box periodicity is built in at the lowest level of the \nq\ subclasses, which are highly optimized for this use case.
The \aabb\ subclass implements a tree data structure of Axis-Aligned Bounding Boxes (AABBs), a type of \gls{bvh} which greatly accelerates the process of finding particles' neighbors \cite{Anderson2016b,Howard2018}.
A second approach is implemented in the \lc\ subclass \cite{Allen1987}, which uses linked cell lists to find particle neighbors.
Both of these classes can find neighboring particle pairs based on a distance cutoff or a desired number of neighbors, and both were adapted from HOOMD-blue.

\begin{figure}
    \centering
    \includegraphics[width=0.5\textwidth]{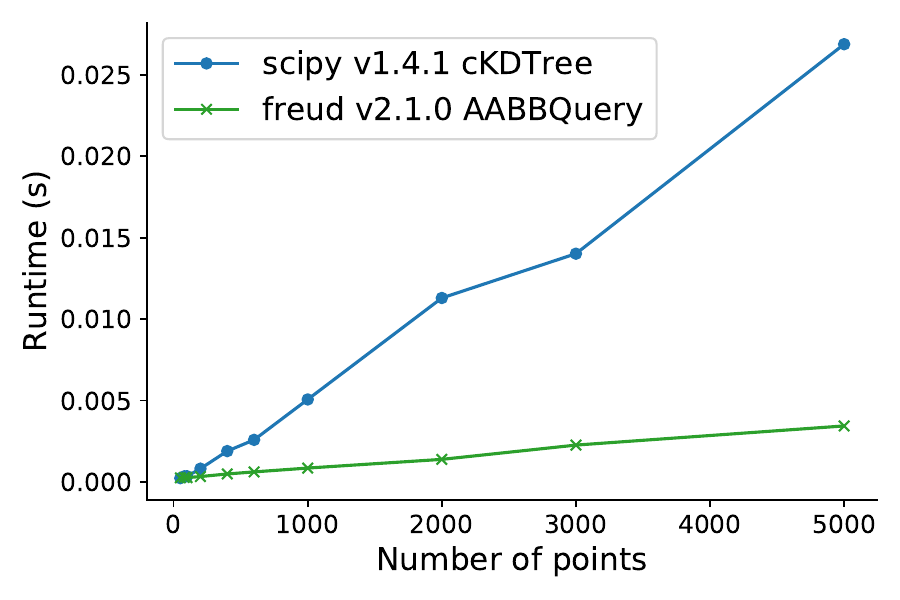}
    \caption{Here we benchmark the \aabb\ implementation in \freud\ against the \texttt{cKDTree} implementation in SciPy.
    We construct randomly generated sets of points such that each particle would have, on average, 12 neighbors within a distance of 1.
    We then measure the performance of finding all neighbors within this distance using both SciPy's \texttt{cKDTree} and \freud's \aabb.
    The benchmarks were performed on a system with an Intel\textsuperscript{\textregistered} Xeon\textsuperscript{\textregistered} CPU E5-2680 v2 @ 2.80GHz.
    The \aabb\ implementation in \freud\ scales much better than SciPy's \texttt{cKDTree} for larger system sizes.
    We do not report error bars due to the extremely low variance in the data.
    The exact details are available at \url{https://github.com/glotzerlab/freud-examples}.
    }
    \label{fig:NeighborPerformance}
\end{figure}

As a performance benchmark, we compare \freud's \aabb\ class with the \texttt{cKDTree} implementation in SciPy \cite{Jones2001}.
As part of SciPy, this implementation is the most readily available alternative to \aabb.
\Cref{fig:NeighborPerformance} shows that \freud's \aabb\ routines clearly outperform the \texttt{cKDTree} as system sizes increase to thousands of points.
Moreover, we note that while \freud\ supports general triclinic boxes, SciPy's \texttt{cKDTree} only supports periodic orthorhombic boxes (i.e. a cuboid, a rectangular prism where all angles are right angles).

In addition to these performance gains, the \nq\ objects in \freud\ are designed to interoperate seamlessly with analysis routines.
Since analyses in \freud\ are written in C++, using a Python \gls{api} to find neighbors and then pass them into other C++ routines would waste time in unnecessary memory transfers.
Furthermore, while a Python \gls{api} should make certain promises, such as sorting the order of the resulting neighbors, the analyses using neighbors simply loop over all pairs and therefore do not require any such extra work.
To avoid this cost, the features of the \nq\ classes are directly accessible in C++ in the form of iterators that lazily produce neighbors.
In practice, using the \nq\ classes in this manner speeds up computations by a factor of two or more depending on the system size.
To make use of these iterators, developers implement analysis methods as lambda functions that are passed as arguments to \freud's internal \gls{tbb} wrappers that apply these functions to neighbor pairs in parallel.

The final feature of the \texttt{locality} module is the \texttt{Voronoi} class, which uses the voro++ library \cite{Rycroft2009} to generate Voronoi diagrams for systems of particles.
Voronoi diagrams are a standard method for characterizing the local geometric arrangements in the system, and they also provide a parameter-free method for defining nearest-neighbor relationships \cite{Lazar2015TopologicalMatter}.
The \texttt{Voronoi} class produces a \nlist\ object that can then be used as the \texttt{neighbors} argument for other compute classes.

\subsection{Analysis Modules}
The remaining modules in \freud\ are independent of one another and contain groups of classes that implement related features.
While some of \freud's features are unique, many others are standard techniques.
However, implementations of these methods commonly lack support for periodicity.
For example, the SciPy library \cite{Jones2001} has functions for computing Voronoi diagrams and correlation functions, but these are restricted to aperiodic systems.

The \texttt{cluster} module of \freud\ can be used to find clusters of particles---where cluster membership is defined by neighbor bonds---and then compute properties of these clusters such as gyration tensors.
The \texttt{density} module contains features for calculating radial distribution functions as well as spatial correlation functions of arbitrary quantities.
Additionally, the \texttt{density} module can estimate local particle density and interpolate particle density onto a regular grid suitable for, e.g., computing discrete Fourier transforms.
The \texttt{interface} module provides a quick tool for identifying interfaces between two mutually exclusive sets of points (e.g. a solid and a liquid phase).
The \texttt{msd} module enables the calculation of mean squared displacements of particles over the course of a trajectory.

\begin{figure*}
    \centering
    \includegraphics[width=0.8\textwidth]{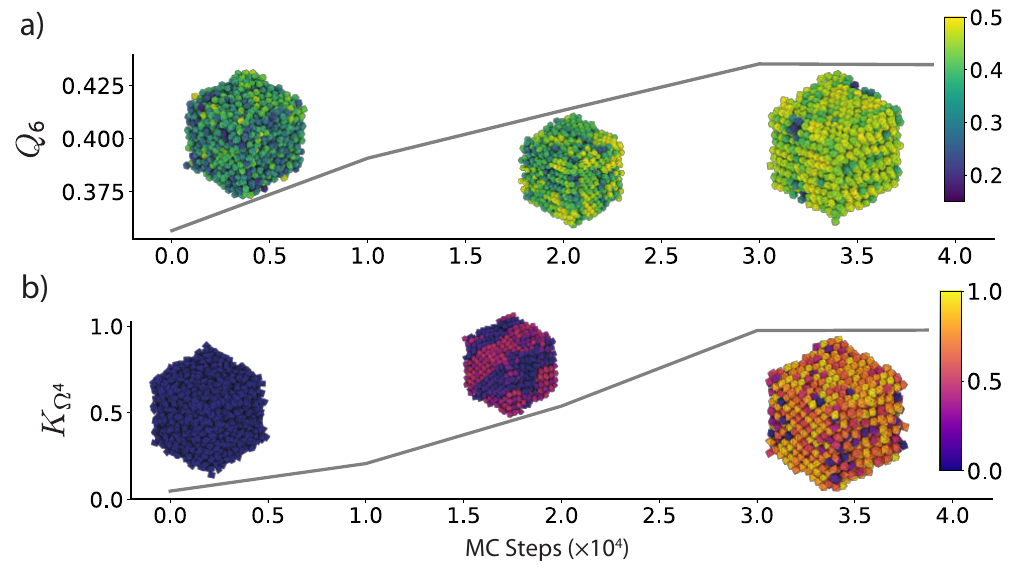}
    \caption{Various order parameters can be used to characterize the degree of ordering in a system. The per-particle order parameter values eventually converge to a uniform global value as the system becomes globally well-ordered. These plots show the evolution of two order parameters over the course of a Monte Carlo simulation of hard particles, which over time rearrange into an ordered phase under compression. Simulation snapshots are colored by the per-particle order parameter and rendered with \texttt{fresnel} \cite{GlotzerLabFresnel}. a) The Steinhardt $Q_6$ order parameter is an appropriate scalar descriptor for systems forming a BCC ($cI2$-W) structure. Systems of cuboctahedra in the fluid phase show a distinctly different characteristic value of the order parameter than in the solid phase. b) The cubatic order parameter $K_{\Omega^4}$ is useful for characterizing ordering in these systems of octahedra.}
    \label{fig:OrderParameters}
\end{figure*}

The \texttt{order} module is the most extensive one in \freud, containing a large number of different order parameters commonly used to measure ordering and identify phase transitions in crystalline systems. 
Of particular note are the bond-orientational order parameters $Q_l$ and $W_l$ \cite{Steinhardt1983} and the cubatic order parameter \cite{Haji-Akbari2015} (see \cref{fig:OrderParameters}).
The module also contains the nematic order parameter for identifying orientationally ordered, translationally disordered phases, as well as a solid-liquid order parameter for identifying generic ordered phases \cite{ReintenWolde1996}.

The other features of \freud\ are analysis methods developed by researchers in our group and not yet implemented anywhere else.
In particular, the \texttt{pmft} and \texttt{environment} modules implement features unique to \freud\ that we now discuss in greater detail.

\subsection{Potentials of Mean Force and Torque}
\label{sec:pmft}
\glsreset{pmft}
The \gls{pmft} is a generalization of the classical \gls{pmf} that was recently developed to quantify directional entropic forces that emerge in crowded systems \cite{VanAnders2014c, VanAnders2014d}.
Given the canonical partition function as a function of particle positions $\{q\}$ and particle orientations $\{Q\}$, ref. \cite{VanAnders2014d} derives the \gls{pmft} by separating out a component corresponding to the relative coordinates of a pair of particles $\Delta \xi_{12}$:
\begin{linenomath*}
    \begin{align}
        Z &= \int d\dchi J(\dchi) e^{-\beta U(\dchi)} \int [d\Tilde{q}][d\Tilde{Q}] e^{-\beta U(\{\Tilde{q}\}, \{\Tilde{Q}\}, \dchi)}  \label{eq:SplitPartition1} \\
        &= \int d\dchi J(\dchi) e^{-\beta U(\dchi)} \,\,\, e^{-\beta \ftilde(\dchi)}
        \label{eq:SplitPartition2}
    \end{align}
\end{linenomath*}
where $J$ is the Jacobian transforming to the local coordinate system and $\ftilde$ is the free energy of the other particles, which have been integrated over in \cref{eq:SplitPartition2}.
The \gls{pmft} $F_{12}$ is defined by the relation
\begin{linenomath*}
    \begin{equation}
        Z \equiv \int d\dchi e^{-\beta F_{12}(\Delta \xi_{12})}
        \label{eq:PMFT}
    \end{equation}
\end{linenomath*}
Combining \cref{eq:SplitPartition2,eq:PMFT} gives an expression for the \gls{pmft}
\begin{linenomath*}
    \begin{equation}
        \beta F_{12}(\dchi) = \beta U(\Delta \xi_{12}) - \log J(\Delta \xi_{12}) + \beta \ftilde(\dchi)
    \end{equation}
\end{linenomath*}
In hard particle systems governed exclusively by excluded volume interactions, the potential energy term becomes an infinite Heaviside function $H$ and the \gls{pmft} can be simplified to
\begin{linenomath*}
    \begin{equation}
        F_{12}(\dchi) = -k_B T \log (H(d(\dchi)) J(\dchi)) + \ftilde(\dchi)
    \end{equation}
\end{linenomath*}

\begin{figure*}[ht]
    \centering
    \includegraphics[width=0.9\textwidth]{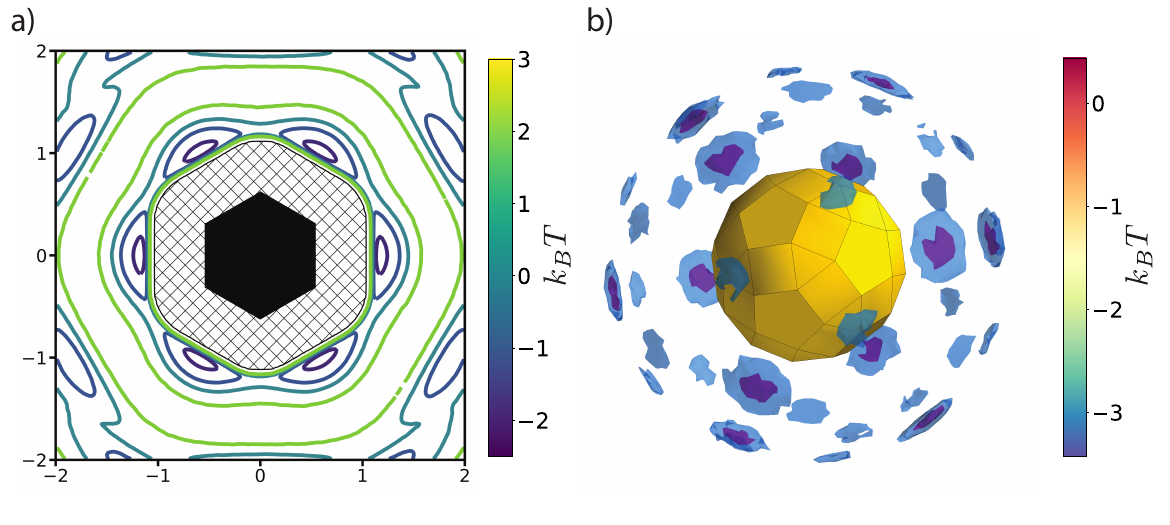}
    \caption{The \gls{pmft} is related to the probability of finding particles at a given position and orientation relative to one another. a) The \gls{pmft} of an ordered system of hexagons \cite{Harper2015a}, where the locations of the wells indicate that particles are much more likely to sit next to the edges of their neighbors than the corners. In two-dimensional systems, the full \gls{pmft} is 3-dimensional, since it also must account for the orientation of the second particle relative to the first; for clarity, in this figure we have integrated out that degree of freedom. b) A \gls{pmft} computed from a system of rhombicosidodecahedra shows distributions of neighboring particles in three dimensions (figure rendered using Mayavi \cite{Ramachandran2011}). There are six degrees of freedom in 3D systems, three translational and three rotational. This \gls{pmft} only shows the three translational degrees of freedom. The wells representing the deepest energy isosurfaces of the \gls{pmft} align with the largest (pentagonal) facets of the polyhedron.}
    \label{fig:PMFT}
\end{figure*}

To contextualize the \gls{pmft}, we note that if in \cref{eq:SplitPartition1} we redefine $\Delta \xi_{12}$ to only include the center-to-center distance of the pair of particles and otherwise follow the same steps, the resulting potential $F_{12}$ reduces to the classical \gls{pmf} with the usual \gls{rdf} relation $g(r) = e^{-\beta F_{12}(r)}$.
This suggests that although the \gls{pmft} is a function of all degrees of freedom required to characterize the relative configuration of a pair of particles, examining a more limited coordinate system can still be informative.
\Cref{fig:PMFT} shows two examples of \glspl{pmft} that contain more information than a \gls{pmf} without containing all available degrees of freedom.
In the 2D \gls{pmft} in the left panel, the orientation of the second particle is ignored, but its angular position relative to the reference particle is sufficient to illustrate the clear preference for facet-to-facet alignment.
Similarly, the right panel ignores the orientation of the second particle (which encodes three degrees of freedom in 3D, as represented by e.g. Euler angles), but once again the preference for facet-to-facet alignment is clear.
For an example of a case where analyzing the full, high-dimensional \gls{pmft} is necessary, see ref. \cite{Harper2019TheCrystals}.

\begin{figure}
    \centering
    \includegraphics[width=0.4\textwidth]{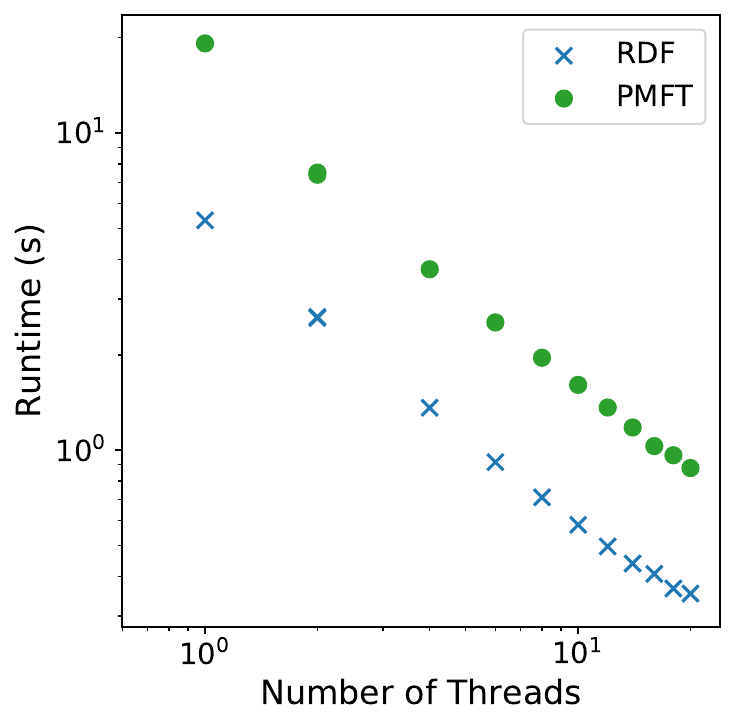}
    \caption{This benchmark compares the performance of the 2D \gls{pmft} to that of an \gls{rdf} on the same two-dimensional system of hexagons used in \cref{fig:PMFT}.
    This computation was performed with a randomly generated trajectory of consisting of 10 frames of 20000 particles.
    Particle positions were constructed such that each particle would have, on average, 12 neighbors within a distance of 1.
    The benchmarks were performed on a system with an Intel\textsuperscript{\textregistered} Xeon\textsuperscript{\textregistered} CPU E5-2680 v2 @ 2.80GHz.
    Both methods have essentially the same performance characteristics, with the \gls{pmft} approximately three times slower than the \gls{rdf}.
    We do not report error bars due to the extremely low variance in the data.}
    \label{fig:PMFTPerformance}
\end{figure}

\freud\ calculates the \gls{pmft} by accumulating a histogram of the configurations of all other particles and then taking the negative logarithm of the counts.
\glspl{pmft} may be accumulated over many frames to generate smoother energy surfaces.
This method of computing the \gls{pmft} is very similar to that of computing an \gls{rdf}, so we compare their scaling behavior in a two-dimensional system in \cref{fig:PMFTPerformance}.
The calculations scale almost identically to many threads, with a constant scaling factor between them.
There are two components contributing to the absolute difference in their performance: 1) the extra operations required to compute the orientation of the local coordinate system in the \gls{pmft}, and 2) the extra cost of binning in multiple dimensions.

We also tested performance as a function of the parameters of these two methods, namely the number of bins and the maximum interparticle distance.
In the latter case, both methods show the expected quadratic growth, since the number of particles included in the calculation increases as the square of this cutoff distance.
The behavior with respect to the number of bins is more interesting: this parameter has no effect on performance until it becomes sufficiently large, at which point performance begins to degrade.
This degradation can be understood as the result of two things: 1) poor cache performance as the histograms become too large to fit in memory, and 2) increasing costs of reduction, which can eventually affect performance.
Since the \gls{pmft} shown is a two-dimensional histogram, the number of bins that can be used along each dimension before experiencing this performance drop is commensurately smaller than can be used with the \gls{rdf}; this effect would be even more pronounced for a three-dimensional \gls{pmft}.

\subsection{Local Environments}
The \texttt{environment} module provides methods for characterizing the local environments of particles that we now illustrate in greater detail.

\subsubsection{Bond-Orientational Order Diagrams}
The \texttt{BondOrder} class enables the calculation of \glspl{bood} \cite{Damasceno2012,Dzugutov1993,Roth1995,Roth2000b,Engel2015}.
Inspired by the bond-orientational order parameters defined by Steinhardt et al. \cite{Steinhardt1983}, \glspl{bood} characterize the local ordering of systems by calculating the vectors between all neighboring particles in a system and then projecting these vectors onto a sphere.
One example of how \glspl{bood} can be used is to identify $n$-fold ordering in a system; in simple crystal structures with $n$-fold coordination, the \gls{bood} will show $n$ peaks corresponding to the average location of nearest neighbors.

In addition to the standard \gls{bood} calculation, the class offers some additional modes of operation that can be useful in specific cases.
One mode involves finding the positions of nearest neighbors in the local coordinate system of a given particle rather than the global coordinate system, which can prevent misidentifying systems with multiple grains \cite{Damasceno2012}.
Another mode modifies the \gls{bood} to help identify plastic crystals, which appear crystalline due to having translational order but lack orientational ordering.
In this mode, the positions of the nearest neighbors of each particle are modified by the relative orientations of the neighboring particles, creating a \gls{bood} in which positional ordering will no longer appear except when orientational ordering is also present.

\subsubsection{Spherical Harmonic Descriptors}
\begin{figure*}
    \centering
    \includegraphics[width=0.8\textwidth]{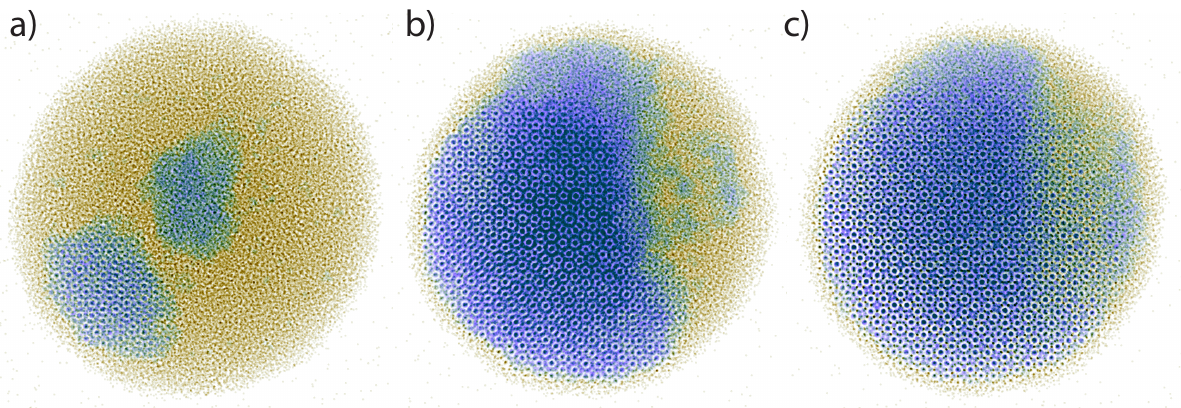}
    \caption{Spherical harmonic descriptors can be used to identify the nucleation and growth of $tP30$-CrFe (Frank-Kasper $\sigma$ phase). a--c) As time progresses, crystallites nucleate and grow. Solid-like particles (blue) are identified via a feedforward artificial neural network using spherical harmonic descriptors (described in more detail in \cite{Spellings2018a}).
    }
    \label{fig:LocalDescriptors}
\end{figure*}

The \gls{bood} is closely related to the Steinhardt order parameters $Q_l$ and $W_l$, which measure rotational order in a system using spherical harmonics \cite{Steinhardt1983}.
While the \gls{bood} is essentially a histogram of nearest-neighbor bonds, the Steinhardt order parameters take this one step further, measuring $l$-fold order by constructing scalar quantities from rotationally invariant combinations of spherical harmonics of degree $l$ calculated from the locations of nearest-neighbor bonds. 
However, spherical harmonic representations can also be used in a variety of different ways.
For example, distinguishing different grains of the same crystal structure could be done using descriptors that are not rotationally invariant.
Alternatively, we can often obtain rotationally-invariant descriptions of local environments for crystal structure identification via the principal axes of the moment of inertia tensor of the environments, or by using particle orientations of anisotropic particles \cite{Spellings2018a,Haji-Akbari2015,Keys2011}.
To support such spherical harmonic analyses, the \texttt{LocalDescriptors} class in \freud\ computes spherical harmonics characterizing particle neighborhoods.
These harmonics can then be combined in arbitrary ways to generate custom descriptors of local particle environments.
Such descriptors have proven useful in identifying multiple complex crystals (see \cref{fig:LocalDescriptors}).
One method for identifying these structures is to use the information contained in this array of spherical harmonics as a set of per-particle features in an artificial neural network (ANN) for structure classification \cite{Spellings2018a}.

\subsubsection{Environment Matching}

Methods like the spherical harmonic descriptors and the \gls{bood} characterize ordering in systems by calculating system-averaged quantities from neighbor bonds.
The \texttt{EnvironmentCluster} class takes a different approach by defining environments according to the nearest neighbors of each particle and performing point set registration to identify and cluster similar environments \cite{Teich2019}.
This type of analysis is particularly useful because it emphasizes local information for each particle.
As a result, it can be used for tasks such as identifying different Wyckoff positions in a crystal.
The complementary \texttt{EnvironmentMotifMatch} class can be used to match clusters to specific motifs, allowing deeper analysis of a given structural motif.

\begin{figure*}
    \centering
    \includegraphics[width=0.9\textwidth]{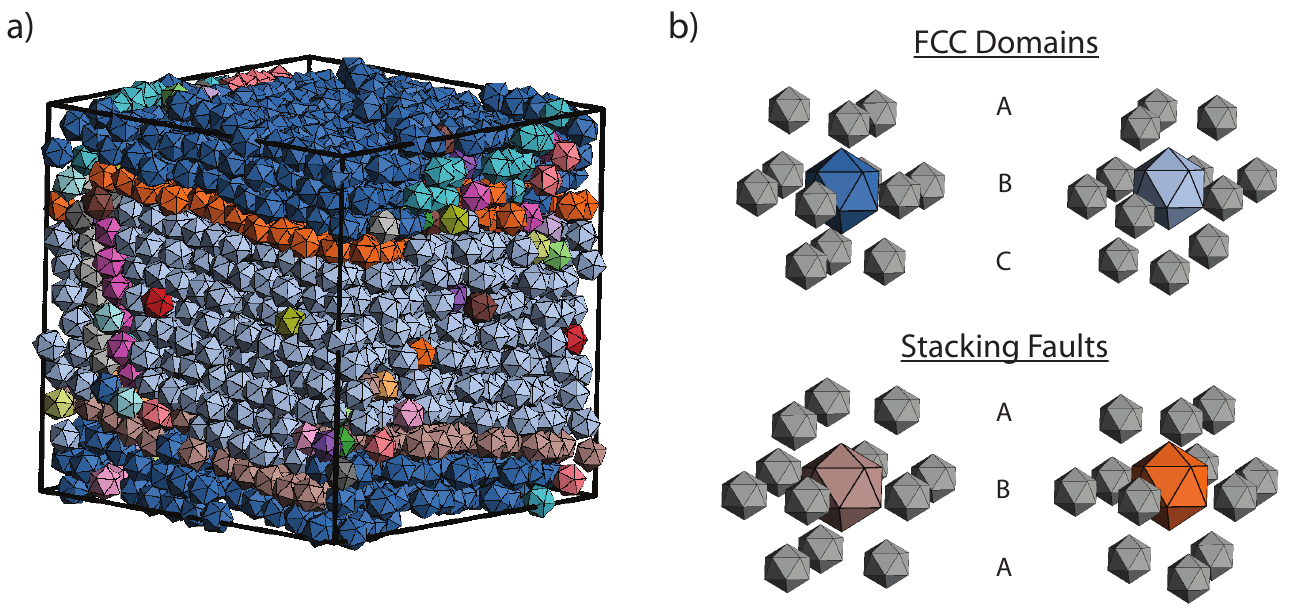}
    \caption{Environment matching allows us to detect variations in the local environments of particles. a) The presence of grain boundaries in this system (rendered with \texttt{fresnel}) is clearly visible due to the different coloring according to local environments. b) The two distinct domains (blue and grey particles) are clustered separately, but we can see that they both exhibit FCC-like ($cF4$-Cu) ordering in their stacking pattern (upper-right). The environment matching method can also detect the different environments of the stacking faults themselves (mauve and orange), which exhibit an ABA stacking pattern instead of the expected ABC pattern (lower-right).}
    \label{fig:EnvMatch}
\end{figure*}

Since this method performs a direct pairwise comparison of all motifs, it is substantially more expensive than common methods used for structure identification.
For instance, the performance is at least an order of magnitude slower than the implementation of Polyhedral Template Matching in OVITO and Common Neighbor Analysis.
Unlike these methods, however, the environment matching algorithm does not depend on previous knowledge of possible structures, and instead infers possible structures entirely from the local motifs present in a system.
Moreover, it can be tuned much more finely than the other methods, allowing not only the identification of crystal structures present, but also the precise identification of stacking faults like those found in \cref{fig:EnvMatch}.
As a result, it is a good complement to existing methods for identifying crystal structures in various systems.

\subsubsection{Angular Separation}
The \texttt{AngularSeparation} classes provide a way to characterize typical particle orientations in a system.
The \texttt{AngularSeparationGlobal} class allows comparison of particle orientations to a set of reference orientations, which can be used to characterize orientational order relative to the reference input.
This metric can be used as an order parameter for measuring orientational disorder in plastic crystals, which exhibit translational order and orientational disorder \cite{Karas2019}.
Alternatively, the \texttt{AngularSeparationNeighbor} computes minimum separation angles between neighboring particles, allowing a more fine-grained analysis of the orientational ordering in local motifs.
Both of these methods account for symmetry by accepting an array of equivalent quaternions corresponding to all symmetry-preserving transformations of the particle (i.e. the particle's point group).

\subsection{Data Generation and Plotting}
For the purposes of teaching via code examples and testing \freud's analyses, \freud\ includes the \texttt{freud.data} module.
It includes the \texttt{UnitCell} class for representing arbitrary unit cells with user-provided box vectors and basis positions.
The \texttt{UnitCell} class includes class methods that generate common crystal structures like face-centered cubic, body-centered cubic, and simple cubic.
The \texttt{data} module also includes a method for generating a random system with uniformly distributed points in a periodic box.

Many analysis modules in \freud\ implement a \texttt{plot()} method, which can be used for rapid data visualization.
The Compute classes (e.g. instances of \texttt{freud.density.RDF}) also define a \texttt{\_repr\_png\_()} method that allows their data to be automatically plotted in IPython environments (such as Jupyter notebooks) using Matplotlib \cite{Hunter2007a}, when the last line in a code cell returns that analysis object.

\section{Examples}
\label{sec:Examples}

In this section, we demonstrate the use of \freud\ in conjunction with the broader scientific software ecosystem.
The code for these examples and many others is available at \url{https://github.com/glotzerlab/freud-examples}.

\subsection{RDF and MSD from LAMMPS simulation}
Here, we consider the problem of calculating the RDF and the MSD of a system simulated using LAMMPS (version 5 Jun 2019) \cite{Plimpton1995a}.
LAMMPS is a standard tool for particle simulation used in many fields, and it supports multiple output formats, including those used by other simulation codes (e.g., the DCD format from CHARMM \cite{Brooks2009} and the XTC format from GROMACS \cite{Berendsen1995a}).
In this case, we demonstrate the case of using the output of a custom dump format in LAMMPS, which allows users to dump selected quantities into a text file.
Although the default XYZ file format lacks sufficient information to calculate an MSD, the necessary particle image information can be included as shown.

\begin{minted}[frame=lines, framesep=2mm, baselinestretch=1.2, bgcolor=LightGray, fontsize=\footnotesize]{python}
import numpy as np
import freud

# For the MSD we also need images, which can be dumped
# using the LAMMPS dump custom command as follows:
# dump 2 all custom 100 output_custom.xyz x y z ix iy iz

# We read the number of particles, the system box, and the
# particle positions into 3 separate arrays.
N = int(np.genfromtxt(
    'output_custom.xyz', skip_header=3, max_rows=1))
box_data = np.genfromtxt(
    'output_custom.xyz', skip_header=5, max_rows=3)
data = np.genfromtxt(
    'output_custom.xyz', skip_header=9, 
    invalid_raise=False)

# Remove the unwanted text rows
data = data[~np.isnan(data).all(axis=1)].reshape(-1, N, 6)

box = freud.box.Box.from_box(
    box_data[:, 1] - box_data[:, 0])
    
# We shift the system by half the box lengths to match the
# freud coordinate system, which is centered at the origin.
# Since all methods support periodicity, this shift is
# simply for consistency but does not affect any analyses.
data[..., :3] -= box.L/2
rdf = freud.density.RDF(bins=100, r_max=4, r_min=1)
for frame in data:
    rdf.compute(system=(box, frame[:, :3]), reset=False)

msd = freud.msd.MSD(box)
msd.compute(
    positions=data[:, :, :3], images=data[:, :, 3:])

# The object contains all the data we need to plot the RDF.
from matplotlib import pyplot as plt
plt.plot(rdf.bin_centers, rdf.rdf)
plt.title('Radial Distribution Function')
plt.xlabel('$r$')
plt.ylabel('$g(r)$')
plt.show()
\end{minted}

If our trajectory was stored in a DCD file, we could modify our code above to read the input data using MDAnalysis (version 0.20.1):

\begin{minted}[frame=lines, framesep=2mm, baselinestretch=1.2, bgcolor=LightGray, fontsize=\footnotesize]{python}
reader = MDAnalysis.coordinates.DCD.DCDReader(
    'output.dcd')
rdf = freud.density.RDF(bins=100, r_max=4, r_min=1)
for frame in reader:
    rdf.compute(system=frame, reset=False)
...
\end{minted}

\subsection{On-the-fly analysis with HOOMD-blue}

A major strength of \freud\ is that it can also be used for on-the-fly analysis.
For example, \freud\ can be used to terminate a simulation based on some additional condition, or to log a quantity at a higher frequency than we want to save the full system trajectory.
In our previous example, we demonstrated the calculation of an RDF using \freud.
An RDF can be noisy when calculated with limited data, so we would like to average it over a large number of simulation frames; however, storing many frames can lead to unreasonably large simulation trajectory files.
Using the simulation engine HOOMD-blue (v2.9.0), we can accumulate RDF data during a simulation without storing the entire output.
Additionally, we show that we can log an order parameter over the course of the simulation:

\begin{minted}[frame=lines, framesep=2mm, baselinestretch=1.2, bgcolor=LightGray, fontsize=\footnotesize]{python}
import hoomd
from hoomd import hpmc
import freud
import numpy as np

hoomd.context.initialize()
system = hoomd.init.create_lattice(
    hoomd.lattice.sc(a=1), n=10)
mc = hpmc.integrate.sphere(seed=42, d=0.1, a=0.1)
mc.shape_param.set('A', diameter=0.5)

rdf = freud.density.RDF(bins=50, r_max=4)
q6 = freud.order.Steinhardt(l=6)

def calc_rdf(timestep):
    snap = system.take_snapshot()
    rdf.compute(system=snap, reset=False)
        
def calc_Q6(timestep):
    snap = system.take_snapshot()
    q6.compute(system=snap,
               neighbors={'num_neighbors': 12})
    return np.mean(q6.particle_order)

# Equilibrate the system before accumulating the RDF.
hoomd.run(1e4)
hoomd.analyze.callback(calc_rdf, period=100)

logger = hoomd.analyze.log(filename='output.log',
                           quantities=['q6'],
                           period=100,
                           header_prefix='#',
                           overwrite=True)

logger.register_callback('q6', calc_Q6)

hoomd.run(1e4)

# Store the computed RDF in a file.
np.savetxt('rdf.csv',
           np.vstack((rdf.bin_centers, rdf.rdf)).T,
           delimiter=',', header='r, g(r)')
\end{minted}

\subsection{Analyzing Atomistic Trajectories from GROMACS}
\label{sec:waterRDF}

As discussed in \cref{sec:Introduction,sec:Design}, \freud's design focus differs from that of many similar tools in the lack of focus on trajectory management.
The example below is based on a simulation trajectory of water molecules in a box generated using GROMACS (version 2020) \cite{Berendsen1995a}.
We use MDTraj (version 1.9.3) \cite{McGibbon2015} to read in an XTC trajectory file and then compute an \gls{rdf} of the oxygen atoms in the water molecules using \freud.
In the process, we demonstrate how the sophisticated subsetting functionality offered by tools like MDTraj can be replicated with Python code, which is very useful when such subsets must be computed from coarse-grained trajectories with highly customized topology definitions that standard trajectory management tools cannot handle.

\begin{minted}[frame=lines, framesep=2mm, baselinestretch=1.2, bgcolor=LightGray, fontsize=\footnotesize]{python}
import mdtraj
import freud
import numpy as np

traj = mdtraj.load_xtc(
    'output/prd.xtc', top='output/prd.gro')
bins = 300
r_max = 1
r_min = 0.01

# Expression selection, a common feature of analysis tools
# for atomistic systems, can be used to identify all
# oxygen atoms.
oxygen_pairs = traj.top.select_pairs('name O', 'name O')

# We can directly use the above selection in freud.
oxygen_indices = traj.top.select('name O')

# Alternatively, we can subset directly using Python logic.
# Such selectors require the user to define the nature of
# the selection, but can be more precisely tailored to a
# specific system.
oxygen_indices = [atom.index for atom in traj.top.atoms
                  if atom.name == 'O']

freud_rdf = freud.density.RDF(
    bins=bins, r_min=r_min, r_max=r_max)
for system in zip(np.asarray(traj.unitcell_vectors),
                  traj.xyz[:, oxygen_indices, :]):
    freud_rdf.compute(system, reset=False)

# We can plot these RDFs to verify that they are equivalent.
from matplotlib import pyplot as plt
fig, ax = plt.subplots()
ax.plot(freud_rdf.bin_centers, freud_rdf.rdf, 'o',
        label='freud', alpha=0.5)
ax.plot(*mdtraj_rdf, 'x', label='mdtraj', alpha=0.5)
ax.set_xlabel('$r$')
ax.set_ylabel('$g(r)$')
ax.set_title('Radial Distribution Function')
ax.legend()
\end{minted}

\subsection{Common Neighbor Analysis}
\label{sec:cna}
\Gls{cna} \cite{Honeycutt1987} is a standard technique for analyzing the local neighborhoods of particles in a crystal.
The method involves a classification of local neighborhoods based on a number of features. Using \freud's \nlist, however, the method is straightforward to implement in Python.

We first consider the simpler problem of identifying all common neighbors between any pair of points.
This is equivalent to searching for the \textbf{second-nearest} neighbor pairs, which can be done using \freud\ as follows (note that this code is primarily written for clarity and could easily be optimized):

\begin{minted}[frame=lines, framesep=2mm, baselinestretch=1.2, bgcolor=LightGray, fontsize=\footnotesize]{python}
import freud
import numpy as np
from collections import defaultdict

# Use a face-centered cubic (fcc) system.
box, points = freud.data.UnitCell.fcc().generate_system(4)
aq = freud.AABBQuery(box, points)
query = aq.query(
    points, {'num_neighbors': 12, 'exclude_ii': True})
nl = query.toNeighborList()

# Get all sets of common neighbors.
common_neighbors = defaultdict(list)
for i, p in enumerate(points):
    selection1 = nl.query_point_indices == i
    for j in nl.point_indices[selection1]:
        selection2 = nl.query_point_indices == j
        for k in nl.point_indices[selection2]:
            if i != k:
                common_neighbors[(i, k)].append(j)
\end{minted}

Our dictionary \texttt{common\_neighbors} now contains lists of common neighbors \texttt{j} for every pair of points \texttt{(i, k)}.
This information could itself be useful for performing some analysis on the system.
If we are interested in actually implementing \gls{cna}, then we need to use this information to build local graphs, which we can do with the \texttt{networkx} (version 2.4) \cite{SciPyProceedings_11} Python package.
Combined with the code above, the \gls{cna} algorithm can be implemented as follows:

\begin{minted}[frame=lines, framesep=2mm, baselinestretch=1.2, bgcolor=LightGray, fontsize=\footnotesize]{python}
import networkx as nx
from collections import Counter

diagrams = defaultdict(list)
particle_counts = defaultdict(Counter)

for (a, b), neighbors in common_neighbors.items():
    # Build up the graph of connections between the
    # common neighbors of a and b.
    g = nx.Graph()
    for i in neighbors:
        for j in set(nl.point_indices[
                nl.query_point_indices == i]
                ).intersection(neighbors):
            g.add_edge(i, j)

    # Define the four identifiers for a CNA diagram:
    #     1. 1 if particles are bonded, 0 if not.
    #     2. Number of shared neighbors.
    #     3. Number of bonds between shared neighbors.
    #     4. Index guaranteeing diagram uniqueness.
    diagram_type = 2 - int(
        b in nl.point_indices[nl.query_point_indices == a])
    key = (diagram_type, len(neighbors),
           g.number_of_edges())
    # If we've seen any neighborhood graphs with this
    # signature, we explicitly check if the two graphs are
    # identical to determine whether to save this one.
    # Otherwise, we always add the new graph.
    if key in diagrams:
        isomorphs = [
            nx.is_isomorphic(g, h) for h in diagrams[key]]
        if any(isomorphs):
            idx = isomorphs.index(True)
        else:
            diagrams[key].append(g)
            idx = diagrams[key].index(g)
    else:
        diagrams[key].append(g)
        idx = diagrams[key].index(g)
    cna_signature = key + (idx,)
    particle_counts[a].update([cna_signature])
\end{minted}

In this code, we are looping over all pairs of previously identified second neighbor shells, and finding bonds between the common neighbors of these pairs.
The graph of these bonds then uniquely identifies a new environment.

\section{Conclusion}
\label{sec:Conclusion}
\freud\ is a high-performance Python library for analyzing particle simulations.
Among simulation analysis packages, \freud\ is unique due to its emphasis on coarse-grained simulations and its flexibility.
Its high-performance C++ back-end makes \freud\ a suitable solution for large-scale, high-throughput simulation analysis, while its simple, compact \gls{api} is highly amenable to integration with other tools for, e.g., machine learning applications.
The package's \gls{api} also promotes the prototyping of new analyses directly in Python, and the intuitive design of \freud's internals ensures that translating such analyses into C++ is a relatively painless process.

The package's design is general enough to work with any particle-based system.
However, \freud\ is primarily targeted at communities of materials scientists, chemical engineers, and physicists analyzing molecular dynamics and Monte Carlo for which existing tools are too specialized to be convenient.
Since it makes no assumptions about the types of its input data or the system topology, \freud\ can be used with arbitrary simulation outputs based on topologies defined by the user.
As a result, \freud\ can find wide use in these areas to simplify workflows that require consideration of periodic systems without the complexity associated with specific atomistic features.
Contributions to this open-source toolkit are highly encouraged as new methods are developed in future research applications.

\section*{Acknowledgments}

Support for the design and development of \freud\ has evolved over time and with programmatic research directions. Conceptualization and early implementations were supported in part by the DOD/ASD(R\&E) under Award No. N00244-09-1-0062 and also by the National Science Foundation, Integrative Graduate Education and Research Traineeship, Award \# DGE 0903629 (E.S.H. and M.P.S.). A majority of the code development including all public code releases was supported by the National Science Foundation, Division of Materials Research under a Computational and Data-Enabled Science \& Engineering Award \# DMR 1409620 (2014-2018) and the Office of Advanced Cyberinfrastructure Award \# OAC 1835612 (2018-2021). V.R. holds the 2019-2020 J. Robert Beyster Computational Innovation Graduate Fellowship at the Uiniversity of Michigan. B.D. acknowledges fellowship support from the National Science Foundation under ACI-1547580, S212: Impl: The Molecular Sciences Software Institute and an earlier National Science Foundation Graduate Research Fellowship Grant \# DGE 1256260 (2016-2019) \cite{Wilkins-Diehr2018,Krylov2018}. M.P.S. also acknowledges support from the University of Michigan Rackham Predoctoral Fellowship program. Computational resources and services supported in part by Advanced Research Computing at the University of Michigan, Ann Arbor.

Declarations of interest: none.

\section*{References}
\bibliographystyle{elsarticle-num-names}
\bibliography{references}

\begin{thebibliography}{66}
\providecommand{\natexlab}[1]{#1}
\providecommand{\url}[1]{\texttt{#1}}
\providecommand{\urlprefix}{URL }
\expandafter\ifx\csname urlstyle\endcsname\relax
  \providecommand{\doi}[1]{doi:\discretionary{}{}{}#1}\else
  \providecommand{\doi}[1]{doi:\discretionary{}{}{}\begingroup
  \urlstyle{rm}\url{#1}\endgroup}\fi
\providecommand{\bibinfo}[2]{#2}

\bibitem[{Anderson et~al.(2017)Anderson, Antonaglia, Millan, Engel, and
  Glotzer}]{Anderson2017b}
\bibinfo{author}{J.~A. Anderson}, \bibinfo{author}{J.~Antonaglia},
  \bibinfo{author}{J.~A. Millan}, \bibinfo{author}{M.~Engel},
  \bibinfo{author}{S.~C. Glotzer}, \bibinfo{journal}{Physical Review X}
  \bibinfo{volume}{7}~(\bibinfo{number}{2}) (\bibinfo{year}{2017})
  \bibinfo{pages}{021001}, \doi{\bibinfo{doi}{10.1103/PhysRevX.7.021001}}.

\bibitem[{Simon et~al.(2019)Simon, Zhou, Ramasubramani, Glaser, Pothukuchy,
  Gollihar, Gerberich, Leggere, Morrow, Jung, Glotzer, Taylor, and
  Ellington}]{Simon2019}
\bibinfo{author}{A.~J. Simon}, \bibinfo{author}{Y.~Zhou},
  \bibinfo{author}{V.~Ramasubramani}, \bibinfo{author}{J.~Glaser},
  \bibinfo{author}{A.~Pothukuchy}, \bibinfo{author}{J.~Gollihar},
  \bibinfo{author}{J.~C. Gerberich}, \bibinfo{author}{J.~C. Leggere},
  \bibinfo{author}{B.~R. Morrow}, \bibinfo{author}{C.~Jung},
  \bibinfo{author}{S.~C. Glotzer}, \bibinfo{author}{D.~W. Taylor},
  \bibinfo{author}{A.~D. Ellington}, \bibinfo{journal}{Nature Chemistry}
  \bibinfo{volume}{11}~(\bibinfo{number}{3}) (\bibinfo{year}{2019})
  \bibinfo{pages}{204--212}, \doi{\bibinfo{doi}{10.1038/s41557-018-0196-3}}.

\bibitem[{Niethammer et~al.(2014)Niethammer, Becker, Bernreuther, Buchholz,
  Eckhardt, Heinecke, Werth, Bungartz, Glass, Hasse, Vrabec, and
  Horsch}]{Niethammer2014}
\bibinfo{author}{C.~Niethammer}, \bibinfo{author}{S.~Becker},
  \bibinfo{author}{M.~Bernreuther}, \bibinfo{author}{M.~Buchholz},
  \bibinfo{author}{W.~Eckhardt}, \bibinfo{author}{A.~Heinecke},
  \bibinfo{author}{S.~Werth}, \bibinfo{author}{H.~J. Bungartz},
  \bibinfo{author}{C.~W. Glass}, \bibinfo{author}{H.~Hasse},
  \bibinfo{author}{J.~Vrabec}, \bibinfo{author}{M.~Horsch},
  \bibinfo{journal}{Journal of Chemical Theory and Computation}
  \bibinfo{volume}{10}~(\bibinfo{number}{10}) (\bibinfo{year}{2014})
  \bibinfo{pages}{4455--4464}, \doi{\bibinfo{doi}{10.1021/ct500169q}}.

\bibitem[{Freddolino et~al.(2008)Freddolino, Liu, Gruebele, and
  Schulten}]{Freddolino2008a}
\bibinfo{author}{P.~L. Freddolino}, \bibinfo{author}{F.~Liu},
  \bibinfo{author}{M.~Gruebele}, \bibinfo{author}{K.~Schulten},
  \bibinfo{journal}{Biophysical Journal}
  \bibinfo{volume}{94}~(\bibinfo{number}{10}) (\bibinfo{year}{2008})
  \bibinfo{pages}{L75--L77}, \doi{\bibinfo{doi}{10.1529/biophysj.108.131565}}.

\bibitem[{Shaw et~al.(2009)Shaw, Bowers, Chow, Eastwood, Ierardi, Klepeis,
  Kuskin, Larson, Lindorff-Larsen, Maragakis, Moraes, Dror, Piana, Shan,
  Towles, Salmon, Grossman, Mackenzie, Bank, Young, Deneroff, and
  Batson}]{Shaw2009}
\bibinfo{author}{D.~E. Shaw}, \bibinfo{author}{K.~J. Bowers},
  \bibinfo{author}{E.~Chow}, \bibinfo{author}{M.~P. Eastwood},
  \bibinfo{author}{D.~J. Ierardi}, \bibinfo{author}{J.~L. Klepeis},
  \bibinfo{author}{J.~S. Kuskin}, \bibinfo{author}{R.~H. Larson},
  \bibinfo{author}{K.~Lindorff-Larsen}, \bibinfo{author}{P.~Maragakis},
  \bibinfo{author}{M.~A. Moraes}, \bibinfo{author}{R.~O. Dror},
  \bibinfo{author}{S.~Piana}, \bibinfo{author}{Y.~Shan},
  \bibinfo{author}{B.~Towles}, \bibinfo{author}{J.~K. Salmon},
  \bibinfo{author}{J.~P. Grossman}, \bibinfo{author}{K.~M. Mackenzie},
  \bibinfo{author}{J.~A. Bank}, \bibinfo{author}{C.~Young},
  \bibinfo{author}{M.~M. Deneroff}, \bibinfo{author}{B.~Batson},
  \bibinfo{title}{{Millisecond-scale molecular dynamics simulations on Anton}},
  in: \bibinfo{booktitle}{Proceedings of the Conference on High Performance
  Computing Networking, Storage and Analysis - SC '09}, \bibinfo{publisher}{ACM
  Press}, \bibinfo{address}{New York, New York, USA}, \bibinfo{pages}{1},
  \doi{\bibinfo{doi}{10.1145/1654059.1654099}}, \bibinfo{year}{2009}.

\bibitem[{McGibbon et~al.(2015)McGibbon, Beauchamp, Harrigan, Klein, Swails,
  Hern{\'{a}}ndez, Schwantes, Wang, Lane, and Pande}]{McGibbon2015}
\bibinfo{author}{R.~T. McGibbon}, \bibinfo{author}{K.~A. Beauchamp},
  \bibinfo{author}{M.~P. Harrigan}, \bibinfo{author}{C.~Klein},
  \bibinfo{author}{J.~M. Swails}, \bibinfo{author}{C.~X. Hern{\'{a}}ndez},
  \bibinfo{author}{C.~R. Schwantes}, \bibinfo{author}{L.-P. Wang},
  \bibinfo{author}{T.~J. Lane}, \bibinfo{author}{V.~S. Pande},
  \bibinfo{journal}{Biophysical Journal}
  \bibinfo{volume}{109}~(\bibinfo{number}{8}) (\bibinfo{year}{2015})
  \bibinfo{pages}{1528--1532}, \doi{\bibinfo{doi}{10.1016/J.BPJ.2015.08.015}}.

\bibitem[{Michaud-Agrawal et~al.(2011)Michaud-Agrawal, Denning, Woolf, and
  Beckstein}]{Michaud-Agrawal2011}
\bibinfo{author}{N.~Michaud-Agrawal}, \bibinfo{author}{E.~J. Denning},
  \bibinfo{author}{T.~B. Woolf}, \bibinfo{author}{O.~Beckstein},
  \bibinfo{journal}{Journal of Computational Chemistry}
  \bibinfo{volume}{32}~(\bibinfo{number}{10}) (\bibinfo{year}{2011})
  \bibinfo{pages}{2319--2327}, \doi{\bibinfo{doi}{10.1002/jcc.21787}}.

\bibitem[{Romo and Grossfield(2009)}]{Romo2009}
\bibinfo{author}{T.~Romo}, \bibinfo{author}{A.~Grossfield},
  \bibinfo{title}{{LOOS: An extensible platform for the structural analysis of
  simulations}}, in: \bibinfo{booktitle}{2009 Annual International Conference
  of the IEEE Engineering in Medicine and Biology Society},
  \bibinfo{publisher}{IEEE}, \bibinfo{pages}{2332--2335},
  \doi{\bibinfo{doi}{10.1109/IEMBS.2009.5335065}}, \bibinfo{year}{2009}.

\bibitem[{Hinsen(2000)}]{Hinsen2000}
\bibinfo{author}{K.~Hinsen}, \bibinfo{journal}{Journal of Computational
  Chemistry} \bibinfo{volume}{21}~(\bibinfo{number}{2}) (\bibinfo{year}{2000})
  \bibinfo{pages}{79--85},
  \doi{\bibinfo{doi}{10.1002/(SICI)1096-987X(20000130)21:2<79::AID-JCC1>3.0.CO;2-B}}.

\bibitem[{Humphrey et~al.(1996)Humphrey, Dalke, and Schulten}]{HUMP96}
\bibinfo{author}{W.~Humphrey}, \bibinfo{author}{A.~Dalke},
  \bibinfo{author}{K.~Schulten}, \bibinfo{journal}{Journal of Molecular
  Graphics} \bibinfo{volume}{14} (\bibinfo{year}{1996})
  \bibinfo{pages}{33--38}, \doi{\bibinfo{doi}{10.1016/0263-7855(96)00018-5}}.

\bibitem[{Kabsch and Sander(1983)}]{Kabsch1983}
\bibinfo{author}{W.~Kabsch}, \bibinfo{author}{C.~Sander},
  \bibinfo{journal}{Biopolymers} \bibinfo{volume}{22}~(\bibinfo{number}{12})
  (\bibinfo{year}{1983}) \bibinfo{pages}{2577--2637},
  \doi{\bibinfo{doi}{10.1002/bip.360221211}}.

\bibitem[{Reinhart and Panagiotopoulos(2018)}]{Reinhart2018}
\bibinfo{author}{W.~F. Reinhart}, \bibinfo{author}{A.~Z. Panagiotopoulos},
  \bibinfo{journal}{Journal of Chemical Physics}
  \bibinfo{volume}{148}~(\bibinfo{number}{12}) (\bibinfo{year}{2018})
  \bibinfo{pages}{124506}, \doi{\bibinfo{doi}{10.1063/1.5021347}}.

\bibitem[{Howard et~al.(2018)Howard, Reinhart, Sanyal, Shell, Nikoubashman, and
  Panagiotopoulos}]{Howard2018}
\bibinfo{author}{M.~P. Howard}, \bibinfo{author}{W.~F. Reinhart},
  \bibinfo{author}{T.~Sanyal}, \bibinfo{author}{M.~S. Shell},
  \bibinfo{author}{A.~Nikoubashman}, \bibinfo{author}{A.~Z. Panagiotopoulos},
  \bibinfo{journal}{Journal of Chemical Physics}
  \bibinfo{volume}{149}~(\bibinfo{number}{9}) (\bibinfo{year}{2018})
  \bibinfo{pages}{094901}, \doi{\bibinfo{doi}{10.1063/1.5043401}}.

\bibitem[{Spellings and Glotzer(2018)}]{Spellings2018a}
\bibinfo{author}{M.~Spellings}, \bibinfo{author}{S.~C. Glotzer},
  \bibinfo{journal}{AIChE Journal} \bibinfo{volume}{64}~(\bibinfo{number}{6})
  (\bibinfo{year}{2018}) \bibinfo{pages}{2198--2206},
  \doi{\bibinfo{doi}{10.1002/aic.16157}}.

\bibitem[{Adorf et~al.(2018)Adorf, Antonaglia, Dshemuchadse, and
  Glotzer}]{Adorf2018b}
\bibinfo{author}{C.~S. Adorf}, \bibinfo{author}{J.~Antonaglia},
  \bibinfo{author}{J.~Dshemuchadse}, \bibinfo{author}{S.~C. Glotzer},
  \bibinfo{journal}{Journal of Chemical Physics}
  \bibinfo{volume}{149}~(\bibinfo{number}{20}) (\bibinfo{year}{2018})
  \bibinfo{pages}{204102}, \doi{\bibinfo{doi}{10.1063/1.5063802}}.

\bibitem[{Vansaders et~al.(2018)Vansaders, Dshemuchadse, and
  Glotzer}]{VanSaders2018}
\bibinfo{author}{B.~Vansaders}, \bibinfo{author}{J.~Dshemuchadse},
  \bibinfo{author}{S.~C. Glotzer}, \bibinfo{journal}{Physical Review Materials}
  \bibinfo{volume}{2}~(\bibinfo{number}{6}) (\bibinfo{year}{2018})
  \bibinfo{pages}{063604},
  \doi{\bibinfo{doi}{10.1103/PhysRevMaterials.2.063604}}.

\bibitem[{Karas et~al.(2016)Karas, Glaser, and Glotzer}]{Karas2016a}
\bibinfo{author}{A.~S. Karas}, \bibinfo{author}{J.~Glaser},
  \bibinfo{author}{S.~C. Glotzer}, \bibinfo{journal}{Soft Matter}
  \bibinfo{volume}{12}~(\bibinfo{number}{23}) (\bibinfo{year}{2016})
  \bibinfo{pages}{5199--5204}, \doi{\bibinfo{doi}{10.1039/c6sm00620e}}.

\bibitem[{Antonaglia et~al.(2018)Antonaglia, van Anders, and
  Glotzer}]{Antonaglia2018}
\bibinfo{author}{J.~A. Antonaglia}, \bibinfo{author}{G.~van Anders},
  \bibinfo{author}{S.~C. Glotzer}, \bibinfo{journal}{arXiv preprint
  arXiv:1803.05936} .

\bibitem[{Du et~al.(2016)Du, van Anders, Newman, and Glotzer}]{Du2016a}
\bibinfo{author}{C.~X. Du}, \bibinfo{author}{G.~van Anders},
  \bibinfo{author}{R.~S. Newman}, \bibinfo{author}{S.~C. Glotzer},
  \bibinfo{journal}{Proceedings of the National Academy of Sciences of the
  United States of America} \bibinfo{volume}{114}~(\bibinfo{number}{20})
  (\bibinfo{year}{2016}) \bibinfo{pages}{E3892--E3899},
  \doi{\bibinfo{doi}{10.1073/pnas.1621348114}}.

\bibitem[{Harper et~al.(2015)Harper, Marson, Anderson, Van~Anders, and
  Glotzer}]{Harper2015a}
\bibinfo{author}{E.~S. Harper}, \bibinfo{author}{R.~L. Marson},
  \bibinfo{author}{J.~A. Anderson}, \bibinfo{author}{G.~Van~Anders},
  \bibinfo{author}{S.~C. Glotzer}, \bibinfo{journal}{Soft Matter}
  \bibinfo{volume}{11}~(\bibinfo{number}{37}) (\bibinfo{year}{2015})
  \bibinfo{pages}{7250--7256}, \doi{\bibinfo{doi}{10.1039/c5sm01351h}}.

\bibitem[{Dice et~al.(2019)Dice, Ramasubramani, Harper, Spellings, Anderson,
  and Glotzer}]{Dice2019}
\bibinfo{author}{B.~Dice}, \bibinfo{author}{V.~Ramasubramani},
  \bibinfo{author}{E.~Harper}, \bibinfo{author}{M.~Spellings},
  \bibinfo{author}{J.~Anderson}, \bibinfo{author}{S.~Glotzer},
  \bibinfo{title}{{Analyzing Particle Systems for Machine Learning and Data
  Visualization with freud}}, in: \bibinfo{booktitle}{Proceedings of the 18th
  Python in Science Conference}, \bibinfo{number}{Scipy},
  \bibinfo{pages}{27--33}, \doi{\bibinfo{doi}{10.25080/Majora-7ddc1dd1-004}},
  \bibinfo{year}{2019}.

\bibitem[{Hunter(2007)}]{Hunter2007a}
\bibinfo{author}{J.~D. Hunter}, \bibinfo{journal}{Computing in Science {\&}
  Engineering} \bibinfo{volume}{9}~(\bibinfo{number}{3}) (\bibinfo{year}{2007})
  \bibinfo{pages}{90--95}, \doi{\bibinfo{doi}{10.1109/MCSE.2007.55}}.

\bibitem[{Plimpton(1995)}]{Plimpton1995a}
\bibinfo{author}{S.~Plimpton}, \bibinfo{journal}{Journal of Computational
  Physics} \bibinfo{volume}{117}~(\bibinfo{number}{1}) (\bibinfo{year}{1995})
  \bibinfo{pages}{1--19}, \doi{\bibinfo{doi}{10.1006/JCPH.1995.1039}}.

\bibitem[{Berendsen et~al.(1995)Berendsen, van~der Spoel, and van
  Drunen}]{Berendsen1995a}
\bibinfo{author}{H.~Berendsen}, \bibinfo{author}{D.~van~der Spoel},
  \bibinfo{author}{R.~van Drunen}, \bibinfo{journal}{Computer Physics
  Communications} \bibinfo{volume}{91}~(\bibinfo{number}{1-3})
  (\bibinfo{year}{1995}) \bibinfo{pages}{43--56},
  \doi{\bibinfo{doi}{10.1016/0010-4655(95)00042-E}}.

\bibitem[{Roe and Cheatham(2013)}]{Roe2013}
\bibinfo{author}{D.~R. Roe}, \bibinfo{author}{T.~E. Cheatham},
  \bibinfo{journal}{Journal of Chemical Theory and Computation}
  \bibinfo{volume}{9}~(\bibinfo{number}{7}) (\bibinfo{year}{2013})
  \bibinfo{pages}{3084--3095}, \doi{\bibinfo{doi}{10.1021/ct400341p}}.

\bibitem[{Case et~al.(2005)Case, Cheatham, Darden, Gohlke, Luo, Merz, Onufriev,
  Simmerling, Wang, and Woods}]{Case2005ThePrograms}
\bibinfo{author}{D.~A. Case}, \bibinfo{author}{T.~E. Cheatham},
  \bibinfo{author}{T.~Darden}, \bibinfo{author}{H.~Gohlke},
  \bibinfo{author}{R.~Luo}, \bibinfo{author}{K.~M. Merz},
  \bibinfo{author}{A.~Onufriev}, \bibinfo{author}{C.~Simmerling},
  \bibinfo{author}{B.~Wang}, \bibinfo{author}{R.~J. Woods},
  \bibinfo{journal}{Journal of Computational Chemistry}
  \bibinfo{volume}{26}~(\bibinfo{number}{16}) (\bibinfo{year}{2005})
  \bibinfo{pages}{1668--1688}, \doi{\bibinfo{doi}{10.1002/jcc.20290}}.

\bibitem[{Schr{\"{o}}dinger(2019)}]{PyMOL}
\bibinfo{author}{L.~Schr{\"{o}}dinger}, \bibinfo{title}{{The PyMOL Molecular
  Graphics System, Version 2.3}}, \bibinfo{year}{2019}.

\bibitem[{Stukowski(2010)}]{Stukowski2010a}
\bibinfo{author}{A.~Stukowski}, \bibinfo{journal}{Modelling and Simulation in
  Materials Science and Engineering}
  \bibinfo{volume}{18}~(\bibinfo{number}{1}),
  \doi{\bibinfo{doi}{10.1088/0965-0393/18/1/015012}}.

\bibitem[{Yesylevskyy(2012)}]{Yesylevskyy2012}
\bibinfo{author}{S.~O. Yesylevskyy}, \bibinfo{journal}{Journal of Computational
  Chemistry} \bibinfo{volume}{33}~(\bibinfo{number}{19}) (\bibinfo{year}{2012})
  \bibinfo{pages}{1632--1636}, \doi{\bibinfo{doi}{10.1002/jcc.22989}}.

\bibitem[{Oliphant(2006)}]{Oliphant2006a}
\bibinfo{author}{T.~E. Oliphant}, \bibinfo{title}{{A guide to NumPy}},
  \bibinfo{publisher}{Trelgol Publishing}, \bibinfo{year}{2006}.

\bibitem[{Lab(2020{\natexlab{a}})}]{GlotzerLabGSD}
\bibinfo{author}{G.~Lab}, \bibinfo{title}{{GSD v2.0.0}},
  \urlprefix\url{https://github.com/glotzerlab/gsd},
  \bibinfo{year}{2020}{\natexlab{a}}.

\bibitem[{Lab(2020{\natexlab{b}})}]{GlotzerLabGarnett}
\bibinfo{author}{G.~Lab}, \bibinfo{title}{{garnett v0.6.1}},
  \urlprefix\url{https://github.com/glotzerlab/garnett},
  \bibinfo{year}{2020}{\natexlab{b}}.

\bibitem[{McKinney(2010)}]{Mckinney2010DataPython}
\bibinfo{author}{W.~McKinney}, \bibinfo{title}{{Data Structures for Statistical
  Computing in Python}}, \bibinfo{type}{Tech. Rep.}, \bibinfo{year}{2010}.

\bibitem[{Allen and Tildesley(1987)}]{Allen1987}
\bibinfo{author}{M.~P. Allen}, \bibinfo{author}{D.~J. Tildesley},
  \bibinfo{title}{{Computer simulation of liquids}},
  \bibinfo{publisher}{Clarendon Press}, ISBN \bibinfo{isbn}{9780198556459},
  \bibinfo{year}{1987}.

\bibitem[{Anderson et~al.(2008)Anderson, Lorenz, and Travesset}]{Anderson2008d}
\bibinfo{author}{J.~A. Anderson}, \bibinfo{author}{C.~D. Lorenz},
  \bibinfo{author}{A.~Travesset}, \bibinfo{journal}{Journal of Computational
  Physics} \bibinfo{volume}{227}~(\bibinfo{number}{10}) (\bibinfo{year}{2008})
  \bibinfo{pages}{5342--5359}, \doi{\bibinfo{doi}{10.1016/J.JCP.2008.01.047}}.

\bibitem[{Glaser et~al.(2015)Glaser, Nguyen, Anderson, Lui, Spiga, Millan,
  Morse, and Glotzer}]{Glaser2015g}
\bibinfo{author}{J.~Glaser}, \bibinfo{author}{T.~D. Nguyen},
  \bibinfo{author}{J.~A. Anderson}, \bibinfo{author}{P.~Lui},
  \bibinfo{author}{F.~Spiga}, \bibinfo{author}{J.~A. Millan},
  \bibinfo{author}{D.~C. Morse}, \bibinfo{author}{S.~C. Glotzer},
  \bibinfo{journal}{Computer Physics Communications} \bibinfo{volume}{192}
  (\bibinfo{year}{2015}) \bibinfo{pages}{97--107},
  \doi{\bibinfo{doi}{10.1016/J.CPC.2015.02.028}}.

\bibitem[{Anderson et~al.(2016)Anderson, Eric~Irrgang, and
  Glotzer}]{Anderson2016b}
\bibinfo{author}{J.~A. Anderson}, \bibinfo{author}{M.~Eric~Irrgang},
  \bibinfo{author}{S.~C. Glotzer}, \bibinfo{journal}{Computer Physics
  Communications} \bibinfo{volume}{204} (\bibinfo{year}{2016})
  \bibinfo{pages}{21--30}, \doi{\bibinfo{doi}{10.1016/j.cpc.2016.02.024}}.

\bibitem[{Behnel et~al.(2011)Behnel, Bradshaw, Citro, Dalcin, Seljebotn, and
  Smith}]{Behnel2011a}
\bibinfo{author}{S.~Behnel}, \bibinfo{author}{R.~Bradshaw},
  \bibinfo{author}{C.~Citro}, \bibinfo{author}{L.~Dalcin},
  \bibinfo{author}{D.~S. Seljebotn}, \bibinfo{author}{K.~Smith},
  \bibinfo{journal}{Computing in Science {\&} Engineering}
  \bibinfo{volume}{13}~(\bibinfo{number}{2}) (\bibinfo{year}{2011})
  \bibinfo{pages}{31--39}, \doi{\bibinfo{doi}{10.1109/MCSE.2010.118}}.

\bibitem[{Calandrini et~al.(2011)Calandrini, Pellegrini, Calligari, Hinsen, and
  Kneller}]{Calandrini2011}
\bibinfo{author}{V.~Calandrini}, \bibinfo{author}{E.~Pellegrini},
  \bibinfo{author}{P.~Calligari}, \bibinfo{author}{K.~Hinsen},
  \bibinfo{author}{G.~Kneller}, \bibinfo{journal}{{\'{E}}cole th{\'{e}}matique
  de la Soci{\'{e}}t{\'{e}} Fran{\c{c}}aise de la Neutronique}
  \bibinfo{volume}{12} (\bibinfo{year}{2011}) \bibinfo{pages}{201--232},
  \doi{\bibinfo{doi}{10.1051/sfn/201112010}}.

\bibitem[{Jones et~al.(2001)Jones, Oliphant, Peterson, and
  {others}}]{Jones2001}
\bibinfo{author}{E.~Jones}, \bibinfo{author}{T.~Oliphant},
  \bibinfo{author}{P.~Peterson}, \bibinfo{author}{{others}},
  \bibinfo{title}{{SciPy: Open source scientific tools for Python}},
  \urlprefix\url{https://www.scipy.org/}, \bibinfo{year}{2001}.

\bibitem[{{Intel}(2020)}]{Intel2018}
\bibinfo{author}{{Intel}}, \bibinfo{title}{{Intel Threading Building Blocks}},
  \urlprefix\url{https://github.com/intel/tbb}, \bibinfo{year}{2020}.

\bibitem[{202(2020)}]{2020AnacondaDistribution}
\bibinfo{title}{{Anaconda Software Distribution}},
  \urlprefix\url{https://anaconda.com}, \bibinfo{year}{2020}.

\bibitem[{{Glotzer Lab}(2020{\natexlab{a}})}]{GlotzerLabFreud}
\bibinfo{author}{{Glotzer Lab}}, \bibinfo{title}{{freud Source Code
  Repository}}, \urlprefix\url{https://github.com/glotzerlab/freud},
  \bibinfo{year}{2020}{\natexlab{a}}.

\bibitem[{Rycroft(2009)}]{Rycroft2009}
\bibinfo{author}{C.~Rycroft}, \bibinfo{title}{{Voro++: a three-dimensional
  Voronoi cell library in C++}}, \bibinfo{type}{Tech. Rep.},
  \bibinfo{institution}{Lawrence Berkeley National Laboratory (LBNL)},
  \bibinfo{address}{Berkeley, CA}, \doi{\bibinfo{doi}{10.2172/946741}},
  \bibinfo{year}{2009}.

\bibitem[{Lazar et~al.(2015)Lazar, Han, and
  Srolovitz}]{Lazar2015TopologicalMatter}
\bibinfo{author}{E.~A. Lazar}, \bibinfo{author}{J.~Han}, \bibinfo{author}{D.~J.
  Srolovitz}, \bibinfo{journal}{Proceedings of the National Academy of Sciences
  of the United States of America} \bibinfo{volume}{112}~(\bibinfo{number}{43})
  (\bibinfo{year}{2015}) \bibinfo{pages}{E5769--E5776},
  \doi{\bibinfo{doi}{10.1073/pnas.1505788112}}.

\bibitem[{{Glotzer Lab}(2020{\natexlab{b}})}]{GlotzerLabFresnel}
\bibinfo{author}{{Glotzer Lab}}, \bibinfo{title}{{fresnel v0.11.0}},
  \urlprefix\url{https://github.com/glotzerlab/fresnel},
  \bibinfo{year}{2020}{\natexlab{b}}.

\bibitem[{Steinhardt et~al.(1983)Steinhardt, Nelson, and
  Ronchetti}]{Steinhardt1983}
\bibinfo{author}{P.~J. Steinhardt}, \bibinfo{author}{D.~R. Nelson},
  \bibinfo{author}{M.~Ronchetti}, \bibinfo{journal}{Physical Review B}
  \bibinfo{volume}{28}~(\bibinfo{number}{2}) (\bibinfo{year}{1983})
  \bibinfo{pages}{784--805}, \doi{\bibinfo{doi}{10.1103/PhysRevB.28.784}}.

\bibitem[{Haji-Akbari and Glotzer(2015)}]{Haji-Akbari2015}
\bibinfo{author}{A.~Haji-Akbari}, \bibinfo{author}{S.~C. Glotzer},
  \bibinfo{journal}{Journal of Physics A: Mathematical and Theoretical}
  \bibinfo{volume}{48}~(\bibinfo{number}{48}) (\bibinfo{year}{2015})
  \bibinfo{pages}{485201}, \doi{\bibinfo{doi}{10.1088/1751-8113/48/48/485201}}.

\bibitem[{Rein~ten Wolde et~al.(1996)Rein~ten Wolde, Ruiz‐Montero, and
  Frenkel}]{ReintenWolde1996}
\bibinfo{author}{P.~Rein~ten Wolde}, \bibinfo{author}{M.~J. Ruiz‐Montero},
  \bibinfo{author}{D.~Frenkel}, \bibinfo{journal}{The Journal of Chemical
  Physics} \bibinfo{volume}{104}~(\bibinfo{number}{24}) (\bibinfo{year}{1996})
  \bibinfo{pages}{9932--9947}, \doi{\bibinfo{doi}{10.1063/1.471721}}.

\bibitem[{van Anders et~al.(2014{\natexlab{a}})van Anders, Ahmed, Smith, Engel,
  and Glotzer}]{VanAnders2014c}
\bibinfo{author}{G.~van Anders}, \bibinfo{author}{N.~K. Ahmed},
  \bibinfo{author}{R.~Smith}, \bibinfo{author}{M.~Engel},
  \bibinfo{author}{S.~C. Glotzer}, \bibinfo{journal}{ACS Nano}
  \bibinfo{volume}{8}~(\bibinfo{number}{1})
  (\bibinfo{year}{2014}{\natexlab{a}}) \bibinfo{pages}{931--940},
  \doi{\bibinfo{doi}{10.1021/nn4057353}}.

\bibitem[{van Anders et~al.(2014{\natexlab{b}})van Anders, Klotsa, Ahmed,
  Engel, and Glotzer}]{VanAnders2014d}
\bibinfo{author}{G.~van Anders}, \bibinfo{author}{D.~Klotsa},
  \bibinfo{author}{N.~K. Ahmed}, \bibinfo{author}{M.~Engel},
  \bibinfo{author}{S.~C. Glotzer}, \bibinfo{journal}{Proceedings of the
  National Academy of Sciences} \bibinfo{volume}{111}~(\bibinfo{number}{45})
  (\bibinfo{year}{2014}{\natexlab{b}}) \bibinfo{pages}{E4812--E4821},
  \doi{\bibinfo{doi}{10.1073/pnas.1418159111}}.

\bibitem[{Ramachandran and Varoquaux(2011)}]{Ramachandran2011}
\bibinfo{author}{P.~Ramachandran}, \bibinfo{author}{G.~Varoquaux},
  \bibinfo{journal}{Computing in Science {\&} Engineering}
  \bibinfo{volume}{13}~(\bibinfo{number}{2}) (\bibinfo{year}{2011})
  \bibinfo{pages}{40--51}, \doi{\bibinfo{doi}{10.1109/MCSE.2011.35}}.

\bibitem[{Harper et~al.(2019)Harper, van Anders, and
  Glotzer}]{Harper2019TheCrystals}
\bibinfo{author}{E.~S. Harper}, \bibinfo{author}{G.~van Anders},
  \bibinfo{author}{S.~C. Glotzer}, \bibinfo{journal}{Proceedings of the
  National Academy of Sciences of the United States of America}
  \bibinfo{volume}{116}~(\bibinfo{number}{34}) (\bibinfo{year}{2019})
  \bibinfo{pages}{16703--16710}, \doi{\bibinfo{doi}{10.1073/pnas.1822092116}}.

\bibitem[{Damasceno et~al.(2012)Damasceno, Engel, and Glotzer}]{Damasceno2012}
\bibinfo{author}{P.~F. Damasceno}, \bibinfo{author}{M.~Engel},
  \bibinfo{author}{S.~C. Glotzer}, \bibinfo{journal}{Science}
  \bibinfo{volume}{337}~(\bibinfo{number}{6093}) (\bibinfo{year}{2012})
  \bibinfo{pages}{453--457}, \doi{\bibinfo{doi}{10.1126/science.1220869}}.

\bibitem[{Dzugutov(1993)}]{Dzugutov1993}
\bibinfo{author}{M.~Dzugutov}, \bibinfo{journal}{Physical Review Letters}
  \bibinfo{volume}{70}~(\bibinfo{number}{19}) (\bibinfo{year}{1993})
  \bibinfo{pages}{2924--2927},
  \doi{\bibinfo{doi}{10.1103/PhysRevLett.70.2924}}.

\bibitem[{Roth et~al.(1995)Roth, Schilling, and Trebin}]{Roth1995}
\bibinfo{author}{J.~W. Roth}, \bibinfo{author}{R.~Schilling},
  \bibinfo{author}{H.~R. Trebin}, \bibinfo{journal}{Physical Review B}
  \bibinfo{volume}{51}~(\bibinfo{number}{22}) (\bibinfo{year}{1995})
  \bibinfo{pages}{15833--15840},
  \doi{\bibinfo{doi}{10.1103/PhysRevB.51.15833}}.

\bibitem[{Roth and Denton(2000)}]{Roth2000b}
\bibinfo{author}{J.~Roth}, \bibinfo{author}{A.~R. Denton},
  \bibinfo{journal}{Physical Review E}
  \bibinfo{volume}{61}~(\bibinfo{number}{6}) (\bibinfo{year}{2000})
  \bibinfo{pages}{6845--6857}, \doi{\bibinfo{doi}{10.1103/PhysRevE.61.6845}}.

\bibitem[{Engel et~al.(2015)Engel, Damasceno, Phillips, and
  Glotzer}]{Engel2015}
\bibinfo{author}{M.~Engel}, \bibinfo{author}{P.~F. Damasceno},
  \bibinfo{author}{C.~L. Phillips}, \bibinfo{author}{S.~C. Glotzer},
  \bibinfo{journal}{Nature Materials}
  \bibinfo{volume}{14}~(\bibinfo{number}{1}) (\bibinfo{year}{2015})
  \bibinfo{pages}{109--16}, \doi{\bibinfo{doi}{10.1038/nmat4152}}.

\bibitem[{Keys et~al.(2011)Keys, Iacovella, and Glotzer}]{Keys2011}
\bibinfo{author}{A.~S. Keys}, \bibinfo{author}{C.~R. Iacovella},
  \bibinfo{author}{S.~C. Glotzer}, \bibinfo{journal}{Journal of Computational
  Physics} \bibinfo{volume}{230}~(\bibinfo{number}{17}) (\bibinfo{year}{2011})
  \bibinfo{pages}{6438--6463}, \doi{\bibinfo{doi}{10.1016/J.JCP.2011.04.017}}.

\bibitem[{Teich et~al.(2019)Teich, van Anders, and Glotzer}]{Teich2019}
\bibinfo{author}{E.~G. Teich}, \bibinfo{author}{G.~van Anders},
  \bibinfo{author}{S.~C. Glotzer}, \bibinfo{journal}{Nature Communications}
  \bibinfo{volume}{10}~(\bibinfo{number}{1}) (\bibinfo{year}{2019})
  \bibinfo{pages}{64}, \doi{\bibinfo{doi}{10.1038/s41467-018-07977-2}}.

\bibitem[{Karas et~al.(2019)Karas, Dshemuchadse, van Anders, and
  Glotzer}]{Karas2019}
\bibinfo{author}{A.~S. Karas}, \bibinfo{author}{J.~Dshemuchadse},
  \bibinfo{author}{G.~van Anders}, \bibinfo{author}{S.~C. Glotzer},
  \bibinfo{journal}{Soft Matter} \doi{\bibinfo{doi}{10.1039/C8SM02643B}}.

\bibitem[{Brooks et~al.(2009)Brooks, Brooks, Mackerell, Nilsson, Petrella,
  Roux, Won, Archontis, Bartels, Boresch, Caflisch, Caves, Cui, Dinner, Feig,
  Fischer, Gao, Hodoscek, Im, Kuczera, Lazaridis, Ma, Ovchinnikov, Paci,
  Pastor, Post, Pu, Schaefer, Tidor, Venable, Woodcock, Wu, Yang, York, and
  Karplus}]{Brooks2009}
\bibinfo{author}{B.~R. Brooks}, \bibinfo{author}{C.~L. Brooks},
  \bibinfo{author}{A.~D. Mackerell}, \bibinfo{author}{L.~Nilsson},
  \bibinfo{author}{R.~J. Petrella}, \bibinfo{author}{B.~Roux},
  \bibinfo{author}{Y.~Won}, \bibinfo{author}{G.~Archontis},
  \bibinfo{author}{C.~Bartels}, \bibinfo{author}{S.~Boresch},
  \bibinfo{author}{A.~Caflisch}, \bibinfo{author}{L.~Caves},
  \bibinfo{author}{Q.~Cui}, \bibinfo{author}{A.~R. Dinner},
  \bibinfo{author}{M.~Feig}, \bibinfo{author}{S.~Fischer},
  \bibinfo{author}{J.~Gao}, \bibinfo{author}{M.~Hodoscek},
  \bibinfo{author}{W.~Im}, \bibinfo{author}{K.~Kuczera},
  \bibinfo{author}{T.~Lazaridis}, \bibinfo{author}{J.~Ma},
  \bibinfo{author}{V.~Ovchinnikov}, \bibinfo{author}{E.~Paci},
  \bibinfo{author}{R.~W. Pastor}, \bibinfo{author}{C.~B. Post},
  \bibinfo{author}{J.~Z. Pu}, \bibinfo{author}{M.~Schaefer},
  \bibinfo{author}{B.~Tidor}, \bibinfo{author}{R.~M. Venable},
  \bibinfo{author}{H.~L. Woodcock}, \bibinfo{author}{X.~Wu},
  \bibinfo{author}{W.~Yang}, \bibinfo{author}{D.~M. York},
  \bibinfo{author}{M.~Karplus}, \bibinfo{journal}{Journal of Computational
  Chemistry} \bibinfo{volume}{30}~(\bibinfo{number}{10}) (\bibinfo{year}{2009})
  \bibinfo{pages}{1545--1614}, \doi{\bibinfo{doi}{10.1002/jcc.21287}}.

\bibitem[{Honeycutt and Andersen(1987)}]{Honeycutt1987}
\bibinfo{author}{J.~D. Honeycutt}, \bibinfo{author}{H.~C. Andersen},
  \bibinfo{journal}{The Journal of Physical Chemistry}
  \bibinfo{volume}{91}~(\bibinfo{number}{19}) (\bibinfo{year}{1987})
  \bibinfo{pages}{4950--4963}, \doi{\bibinfo{doi}{10.1021/j100303a014}}.

\bibitem[{Hagberg et~al.(2008)Hagberg, Schult, and Swart}]{SciPyProceedings_11}
\bibinfo{author}{A.~A. Hagberg}, \bibinfo{author}{D.~A. Schult},
  \bibinfo{author}{P.~J. Swart}, \bibinfo{title}{{Exploring Network Structure,
  Dynamics, and Function using NetworkX}}, in: \bibinfo{editor}{G.~Varoquaux},
  \bibinfo{editor}{T.~Vaught}, \bibinfo{editor}{J.~Millman} (Eds.),
  \bibinfo{booktitle}{Proceedings of the 7th Python in Science Conference},
  \bibinfo{address}{Pasadena, CA USA}, \bibinfo{pages}{11--15},
  \bibinfo{year}{2008}.

\bibitem[{Wilkins-Diehr and Crawford(2018)}]{Wilkins-Diehr2018}
\bibinfo{author}{N.~Wilkins-Diehr}, \bibinfo{author}{D.~T. Crawford},
  \bibinfo{journal}{Computing in Science and Engineering}
  \bibinfo{volume}{20}~(\bibinfo{number}{5}) (\bibinfo{year}{2018})
  \bibinfo{pages}{26--38}, \doi{\bibinfo{doi}{10.1109/MCSE.2018.05329813}}.

\bibitem[{Krylov et~al.(2018)Krylov, Windus, Barnes, Marin-Rimoldi, Nash,
  Pritchard, Smith, Altarawy, Saxe, Clementi, Crawford, Harrison, Jha, Pande,
  and Head-Gordon}]{Krylov2018}
\bibinfo{author}{A.~Krylov}, \bibinfo{author}{T.~L. Windus},
  \bibinfo{author}{T.~Barnes}, \bibinfo{author}{E.~Marin-Rimoldi},
  \bibinfo{author}{J.~A. Nash}, \bibinfo{author}{B.~Pritchard},
  \bibinfo{author}{D.~G. Smith}, \bibinfo{author}{D.~Altarawy},
  \bibinfo{author}{P.~Saxe}, \bibinfo{author}{C.~Clementi},
  \bibinfo{author}{T.~D. Crawford}, \bibinfo{author}{R.~J. Harrison},
  \bibinfo{author}{S.~Jha}, \bibinfo{author}{V.~S. Pande},
  \bibinfo{author}{T.~Head-Gordon}, \bibinfo{journal}{Journal of Chemical
  Physics} \bibinfo{volume}{149}~(\bibinfo{number}{18}) (\bibinfo{year}{2018})
  \bibinfo{pages}{180901}, \doi{\bibinfo{doi}{10.1063/1.5052551}}.

\end{thebibliography}
\end{document}